\documentclass[10pt,twocolumn,twoside]{IEEEtran}

\ifCLASSINFOpdf
\else
\fi

\usepackage{amsmath,graphicx,subfigure,cite}

\usepackage[T1]{fontenc}
\makeindex

\begin{document}
\title{An Efficient Parameterization of the Room Transfer Function}
\author{Prasanga~Samarasinghe*,~\IEEEmembership{Student Member,~IEEE,} 
        Thushara~Abhayapala,~\IEEEmembership{Senior Member,~IEEE,}
        Mark~Poletti,~\IEEEmembership{Senior Member,~IEEE,}
        and~Terence~Betlehem,~\IEEEmembership{Senior Member,~IEEE}}

\maketitle
\begin{abstract}
This paper proposes an efficient parameterization of the Room Transfer Function (RTF). Typically, the RTF rapidly varies with varying source and receiver positions, hence requires an impractical  number of point to point measurements to characterize a given room. Therefore, we derive a novel RTF parameterization that is robust to both receiver and source variations with the following salient features: (i) The parameterization is given in terms of a modal expansion of $3$D basis functions.   (ii)  The aforementioned modal expansion  can be truncated at a finite number of modes given that the source and receiver locations are from two sizeable spatial regions, which are arbitrarily distributed.  (iii)  The parameter weights/coefficients are independent of the source/receiver positions. Therefore, a finite set of  coefficients is shown to be capable of accurately calculating the RTF between any two arbitrary points from a pre-defined spatial region where the source(s) lie and a pre-defined spatial region where the receiver(s) lie.  A practical method to measure the  RTF coefficients is also provided, which only requires a single microphone unit and a single loudspeaker unit, given that the room characteristics remain stationary over time. The accuracy of the above parameterization is verified using appropriate simulation examples.
\end{abstract}

\IEEEpeerreviewmaketitle

\section{Introduction}
\label{sec:Intro}

The room transfer function (RTF), demonstrates the collective effect of multipath propagation of sound between a source and a receiver within a given room enclosure. Accurate modeling of the RTF  is useful in soundfield simulators as well as many other applications such as sound reproduction, soundfield equalization, echo cancellation, and speech dereverberation. These applications use appropriate RTF deconvolution methods to cancel the effects of room reflections (reverberation), and therefore, are highly dependent on the accuracy of the RTF model.

The theoretical  solution to the RTF based on the Green's function\cite{kuttruff2009room} was derived assuming a strict rectangular room geometry.  It can only be applied to highly idealised cases with reasonable effort. The rooms with which we are concerned in our daily life however are more or less irregular in shape and the formulation of irregular boundary conditions will require extensive numerical calculations. For this reason, the immediate application of the classical model to practical problems in room acoustics is limited. 

In practice, RTFs are usually estimated as FIR filters, or as   parametric equations based on the geometrical properties of the room.  In the FIR filter approach, the RTF is assumed to behave as a linear time-invariant system, and then   modeled as either  an all-zero, all-pole, pole-zero \cite{mourjopoulos1991pole} or a common  pole-zero \cite{haneda1994common} system. The coefficients of these models are estimated as variable parameters of the RTF, and since the  RTF is extremely sensitive to source and receiver variations, the coefficients too experience a similar sensitivity  \cite{haneda1994common}. This problem not only requires  repetitive parameter calculations with varying source/receiver positions, but also demands for adaptive inverse-filters with cumbersome processing algorithms during equalization \cite{radlovic2000equalization, hikichi2007inverse}.     Furthermore, in practice, the time invariant aspect of  room acoustics is far from reality \cite{hikichi2007inverse}, which remains as a fundamental weakness of the time-invariant filter model. 

In contrast, the  geometric room acoustics model,  heavily relies on the room geometry and ray optic methods borrowed from computer graphics.  The first geometric model for room reverberation was introduced by  Allen and Berkley \cite{allen1979image}. This work became the basis for many subsequent geometric models and is based on the notion that  reverberation can be represented as the effect of an infinite number of image sources that are created by reflecting the true acoustic source in room walls. A faster algorithm to evaluate the image source method for single source-multiple receiver applications was later introduced in \cite{duraiswami2007fast} using the multipole expansion. Other common geometric models include ray tracing \cite{krokstad1968calculating}, beam tracing  \cite{funkhouser2004beam}, acoustic radiosity  \cite{hodgson2006experimental}, and Finite Difference Time Domain (FDTD) \cite{botteldooren1995finite, kowalczyk2011room} methods.  Even though these techniques have certain similarities, their theoretical foundations are often unique for each method. For example, the  ray tracing method assumes high operating frequencies while the FDTD method assumes a low-mid frequency bandwidth \cite{botteldooren1995finite}\footnote{At high frequencies, the computational cost is too high due to the increased number of points (small wavelengths).}.  Therefore, their applicability to a general room is quite limited.  More generalized geometric models incorporating multiple specialized models were recently introduced in \cite{siltanen2007room,southern2011spatial, lehmann2010diffuse}. However, due to the lack of preciseness in reflection methods, and the vast variation of room geometries available,  an exact estimation of the RTF based on geometrical properties remain unresolved. 

Due to the inefficiency of  existing RTF models,   alternative  equalization techniques tend to measure the RTF at a finite set of points which are later incorporated to the sound processing algorithm directly \cite{mourjopoulos1994digital, bharitkar2001cluster,tervo2013spatial}.  However, as explained earlier,  even a  small-scale variation in source/receiver positions   results in a drastic variation in the  RTF   \cite{radlovic1999poor}, and therefore, the above method only  gives accurate results at the design points, while the performance degradation present elsewhere is too significant. Additional limitations are caused by the inaccuracies  involved with the point-point RTF measurements. Recent work on improving the RTF measurement via modified source and receiver directivity patterns include \cite{farina2007measuring, pollow2013including, pelzercontinuous, farina2006room}.

A complete equalization solution that is robust to receiver point variations was first proposed in \cite{betlehem2005theory} for $2$D applications,  which exploits a novel RTF model based on the harmonic solution to the wave equation. This model parameterizes the RTF between a fixed source and any arbitrary point within a source-free receiver region in terms of a weighted sum of $2$D basis functions, while the weights need to be separately measured. Thus, the successful extraction of a finite set of parameter weights/coefficients enables RTF characterization between a given source location and any arbitrary point within a given receiver region.  However, these coefficients remain unique to the source location of interest, and therefore, the slightest variation in source positioning requires a new set of RTF parameters to be measured.

In this paper, we introduce an efficient RTF parameterization in $3$D,  that is robust to both receiver and source variations so that the extraction of a finite set of coefficients is sufficient to characterize an entire room enclosure of interest. In other words, we derive a  $3$D model, which characterizes the RTF between any two arbitrary points from a primary  spatial region where the source(s) lie and a secondary spatial region where the receiver(s) lie.  More importantly, we impose no restrictions  on the geometrical configuration of the source and receiver regions and as a result,  the proposed parameterization is valid for any two arbitrary points from the given room. Following  \cite{betlehem2005theory}, this  parameterization is  based on the harmonic solution to the wave equation, and therefore, is derived in terms of a weighted sum of $3$D  basis functions. Furthermore, it only requires a minimum  of $(N_{s}+1)^2(N_{r}+1)^2$ coefficients to characterize the RTF over  an $N_{s}^{\text{th}}$ order source region   and an $N_{r}^{\text{th}}$ order receiver region\footnote{Section \ref{ssec:modDecomp} discusses how  the order of a spatial soundfield is determined over a known region and a given frequency.}.  We also provide a practical  method to extract the aforementioned coefficients, which only requires RTF measurements over a finite set of source-receiver combinations  and associated numerical processing. Given the room characteristics remain stationary over time, these measurements can be obtained using a single microphone unit and a single loudspeaker unit.

The paper is structured as follows.  In Sec. \ref{sec:Para}, we first decompose the room response into direct and reverberant components where the former is known and the latter is unknown. The unknown reverberant component is then parameterized in terms of a weighted sum of $ 3$D basis functions.  In Sec \ref{sec:Coefs}, we describe a robust method to obtain the parameter weights, which only requires   a finite set of RTF measurements. Finally, in Sec. \ref{sec:Sim}, we demonstrate the accuracy of the proposed parameterization, by comparing it with a simulated room based on the image source model. This section also presents an error analysis performed over a broadband frequency range.

\section{Parameterization  of the room transfer function}
\label{sec:Para}
\subsection{Problem formulation} 
The main objective of this paper is to have an efficient  parameterization for the RTF such that it is valid for  variations in the receiver position as well as in the source position.  Therefore, we first define a continuous spatial region where the source(s) lie {\em (source region)} and a continuous spatial region where the receiver(s) lie {\em (receiver region)}, and the new parameterization is expected to deliver the RTF between any two arbitrary points from these two regions. 

For computational simplicity, we assume the receiver region named $\eta$ to be a sphere of radius $R_{r}$ centered at the origin $\boldsymbol{O}$ and the source region named $\zeta$  to be another sphere of radius $R_{s}$ centered at $\boldsymbol{O_{s}}$ (See Fig.~\ref{fig:config}). In spherical coordinates, the receiver point within $\eta$ is denoted by $\boldsymbol{x}=(x,\theta_{x},\phi_{x})$ and the source location within $\zeta$ is denoted by $\boldsymbol{y}=(y,\theta_{y},\phi_{y})$ where  $\boldsymbol{y}=\boldsymbol{y}^{(s)}+\boldsymbol{R_{sr}}$ with $\boldsymbol{y}^{(s)}=(y^{(s)},\theta^{(s)}_{y},\phi^{(s)}_{y})$ representing the same source location with respect to $\boldsymbol{O_{s}}$  and $\boldsymbol{R_{sr}}$ representing the vector connecting $\boldsymbol{O}$ to $\boldsymbol{O_{s}}$. 

In a reverberant environment, the acoustic transfer function  between $\boldsymbol{x}$ and $\boldsymbol{y}$ can be decomposed in to a direct path field and a reflected field as
\begin{equation} 
\label{eq:P}
H(\boldsymbol{x},\boldsymbol{y},\textit{k})=H_{\text{dir}}(\boldsymbol{x},\boldsymbol{y},\textit{k})+H_{\text{rvb}}(\boldsymbol{x},\boldsymbol{y},\textit{k})
\end{equation}
\noindent where $\textit{k} = 2\pi\textit{f}/\textit{c}$ is the wave number,  \textit{f} is the frequency and \textit{c} is the speed of sound propagation. The direct field component due to a unit amplitude point source at $\boldsymbol{y}$ is independent of the room characteristics and can be given in terms of \cite{poletti2005three}
\begin{equation} 
\label{eq:Hdir}
H_{\text{dir}}(\boldsymbol{x},\boldsymbol{y},\textit{k})=\frac{e^{i\textit{k}\left\|\boldsymbol{x}-\boldsymbol{y}\right\|}}{4\pi\left\|\boldsymbol{x}-\boldsymbol{y}\right\|}.
\end{equation}

However,  $H_{\text{rvb}}(\boldsymbol{x},\boldsymbol{y},\textit{k})$, the corresponding reflected field   incident at $\eta$  is unknown, and completely dependent on the room characteristics. Our aim is to parameterize this unknown field so that a finite set of weights/coefficients unique to the room will be capable of  predicting $H_{\text{rvb}}(\boldsymbol{x},\boldsymbol{y},\textit{k})$  between any two points from $\zeta$ and $\eta$.

We base our parameterization approach on the fact that the unknown $H_{\text{rvb}}(\boldsymbol{x},\boldsymbol{y},\textit{k})$ incident on $\eta$ is  caused by the outward propagating wavefield from $\zeta$. Since both these incoming and outgoing   soundfields can be represented in terms of modal decompositions\footnote{A decomposition using the basis functions of the solution to the wave equation.}, $H_{\text{rvb}}(\boldsymbol{x},\boldsymbol{y},\textit{k})$ could also be represented in terms of a similar decomposition. The  coefficients of such a decomposition  will then enable the user to  predict  the RTF between two arbitrary points from  $\zeta$ and $\eta$.  Following the above concept, we first decompose the reverberant field at $\eta$ due to an arbitrary outgoing field from $\zeta$ and then derive an exact decomposition for the  room transfer function. 

\subsection{Modal decomposition of an arbitrary  reverberant field}
\label{ssec:modDecomp}
Consider an  arbitrary outgoing field from $\zeta$, which can be represented in terms of a spherical harmonic decomposition\footnote{Other coordinate systems could be used instead of spherical coordinates, resulting in a different set of basis functions.}  with respect to $\boldsymbol{O_{s}}$  as
\begin{equation} 
\label{eq:3DExtSound_}
S_{\text{out}}(\boldsymbol{z^{(s)}},\textit{k})=\sum_{n=0}^{N_{s}}\sum_{m=-n}^{n}\beta_{nm}^{(s)}(\textit{k})h_{n}(\textit{k}z^{(s)} )Y_{nm}(\theta^{(s)}_{z},\phi^{(s)}_{z})
\end{equation}

\noindent where $\boldsymbol{z^{(s)}}=(z^{(s)},\theta^{(s)}_{z},\phi^{(s)}_{z})$ denotes the observation point outside of $\zeta$, $\beta_{nm}^{(s)}(\textit{k})$ denotes the coefficients of the outgoing soundfield caused by the source distribution in $\zeta$,  $Y_{nm}(\theta,\phi)$ denotes the spherical harmonic of order $n$ and degree $m$, $h_{n}(\cdot)$ represents the spherical Hankel function of the first kind with  order $n$ and $N_{s}=\left\lceil \textit{k}eR_{s}/2\right\rceil$ denotes the exterior field truncation limit  for a source distribution with its furthest source located at a distance of $R_{s}$ \cite{samarasinghe20123d}\footnote{The truncation of a spherical harmonic based soundfield decomposition was originally derived based on the high pass behavior of Bessel functions. More precisely, Bessel functions of the form $j_{n}(x)$ at $x\leq kr$ tend to be close to zero  for orders above $N=\textit{k}er/2$, and play an insignificant role in the infinite summation. In case the reader is confused by the absence of Bessel functions in (\ref{eq:3DExtSound_}), please note that the modal coefficients $\beta_{nm}^{(s)}(\textit{k})$ of any arbitrary outgoing soundfield can be represented in terms of Bessel functions \cite{samarasinghe20123d}.}.

If the resulting reflected field at $\eta$ due to each unit amplitude outgoing mode of (\ref{eq:3DExtSound_}) can be extracted, the total reflected field caused by an arbitrary outgoing field can be successfully predicted. To demonstrate the above statement, let's consider a unit amplitude outgoing wave of order $n'$ and mode $m'$
\begin{equation}
\label{eq:beta2}
\beta_{nm}^{(s)}(\textit{k})=\begin{cases} 1,& \mbox{ $n=n'$ and $m=m'$ }\\
0,& \mbox{otherwise},
\end{cases}
\end{equation}

\noindent producing 

\begin{equation} 
\label{eq:3DExtSound1_}
S_{\text{out}}(\boldsymbol{z^{(s)}},\textit{k})=h_{n'}(\textit{k}z^{(s)})Y_{n'm'}(\theta^{(s)}_{z},\phi^{(s)}_{z}).
\end{equation}

\noindent For this particular outgoing source field, there will be a resulting reflected field present at the receiver region  $\eta$. Irrespective of the geometrical configuration of $\zeta$ and $\eta$,  the mirror images of the sources within  $\zeta$ are always outside of $\eta$  and therefore, the aforementioned reflected field will be a source free incoming field. Such a soundfield can be given in terms of a harmonic decomposition of the form

\begin{equation} 
\label{eq:3DintSound}
R_{n'm'}(\boldsymbol{x},\textit{k})=\sum_{v=0}^{N_{r}}\sum_{\mu=-v}^{v}\alpha^{n'm'}_{v\mu}(\textit{k})j_{v}(\textit{k}x)Y_{v\mu}(\theta_{x},\phi_{x})
\end{equation}
\noindent where  $\alpha^{n'm'}_{v\mu}(\textit{k})$ denotes the soundfield coefficients of the reverberant  field incident at $\eta$  caused by  an unit amplitude $n'^{\text{th}}$ order and $m'^{\text{th}}$ mode outgoing soundfield at $\zeta$, $j_{n}(\cdot)$ represents the spherical Bessel function of order $n$ and $N_{r}=\left\lceil \textit{k}eR_{r}/2\right\rceil$ denotes the interior field truncation limit \cite{samarasinghe20123d}\footnote{Truncation is derived following the same principle discussed earlier.}. If $\alpha^{n'm'}_{v\mu}(\textit{k})$ of (\ref{eq:3DintSound}) can be recorded up to order $N_{r}$ for each unit amplitude outgoing mode from $\zeta$, the reverberant field at $\eta$ due to an arbitrary outgoing field at $\zeta$ can be derived using (\ref{eq:3DExtSound_}), (\ref{eq:3DExtSound1_}) and (\ref{eq:3DintSound}) as

\begin{equation} 
\label{eq:Prvb1}
\begin{split}
P_{\text{rvb}}(\boldsymbol{x},\textit{k})&=\sum_{n=0}^{N_{s}}\sum_{m=-n}^{n}\sum_{v=0}^{N_{r}}\sum_{\mu=-v}^{v}\beta_{nm}^{(s)}(\textit{k})\alpha^{nm}_{v\mu}(\textit{k})j_{v}(\textit{k}x) \\
&  Y_{v\mu}(\theta_{x},\phi_{x}).
\end{split}
\end{equation}

\subsection{Modal decomposition of the room transfer function}

\noindent Now consider a unit amplitude point source at $\boldsymbol{y^{(s)}} \in \zeta$, producing outgoing soundfield coefficients $\beta_{nm}^{(s)}(\textit{k})$  of the form   \cite{poletti2005three}

\begin{equation} 
\label{eq:beta4}
\beta_{nm}^{(s)}(\textit{k})=i\textit{k}j_{n}(\textit{k}y^{(s)})Y^{*}_{nm}(\theta^{(s)}_{y},\phi^{(s)}_{y}).
\end{equation} 

\noindent The corresponding reflected field at $\eta$ describes the unknown reverberant component $H_{\text{rvb}}(\boldsymbol{x},\boldsymbol{y},\textit{k})$ of (\ref{eq:P}). This can be derived using (\ref{eq:Prvb1}) and  (\ref{eq:beta4}) as

\begin{equation} 
\label{eq:Hrvb}
\begin{split}
H_{\text{rvb}}(\boldsymbol{x},\boldsymbol{y},\textit{k})&=i\textit{k}\sum_{n=0}^{N_{s}}\sum_{m=-n}^{n}\sum_{v=0}^{N_{r}}\sum_{\mu=-v}^{v}\alpha^{nm}_{v\mu}(\textit{k})j_{n}(\textit{k} y^{(s)})j_{v}(\textit{k}x) \\
& Y^{*}_{nm}(\theta^{(s)}_{y},\phi^{(s)}_{y})Y_{v\mu}(\theta_{x},\phi_{x}).
\end{split}
\end{equation}
\noindent Therefore, the total acoustic transfer function between any two arbitrary  points from the source region $\zeta$ and the receiver region $\eta$ can be given in terms of the  direct field (\ref{eq:Hdir}) and reflected field (\ref{eq:Hrvb}) components as 

\begin{equation} 
\label{eq:H}
\begin{split}
H(\boldsymbol{x},\boldsymbol{y},\textit{k})&=\frac{e^{i\textit{k}\left\|\boldsymbol{x}-\boldsymbol{y}\right\|}}{4\pi\left\|\boldsymbol{x}-\boldsymbol{y}\right\|}+i\textit{k}\sum_{n=0}^{N_{s}}\sum_{m=-n}^{n}\sum_{v=0}^{N_{r}}\sum_{\mu=-v}^{v}\alpha^{nm}_{v\mu}(\textit{k}) \\
& j_{n}(\textit{k}y^{(s)})j_{v}(\textit{k}x)Y^{*}_{nm}(\theta^{(s)}_{y},\phi^{(s)}_{y})Y_{v\mu}(\theta_{x},\phi_{x}).
\end{split}
\end{equation}

\begin{figure}%
\centering
\subfigure[Non overlapping]{
\label{fig:one}
\includegraphics[width=0.7\columnwidth]{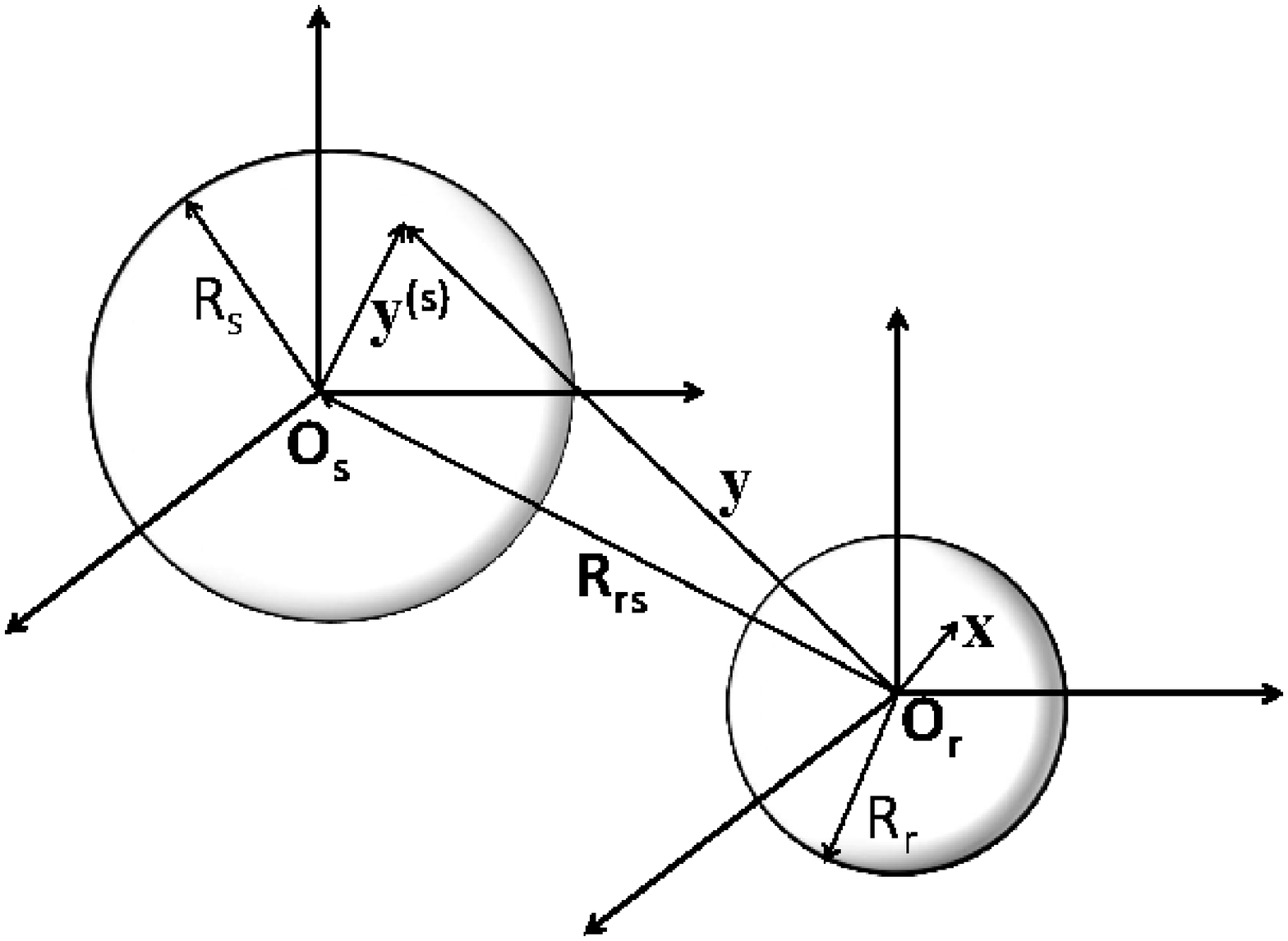}}\\
\subfigure[Concentric]{
\label{fig:two}
\includegraphics[width=0.4\columnwidth]{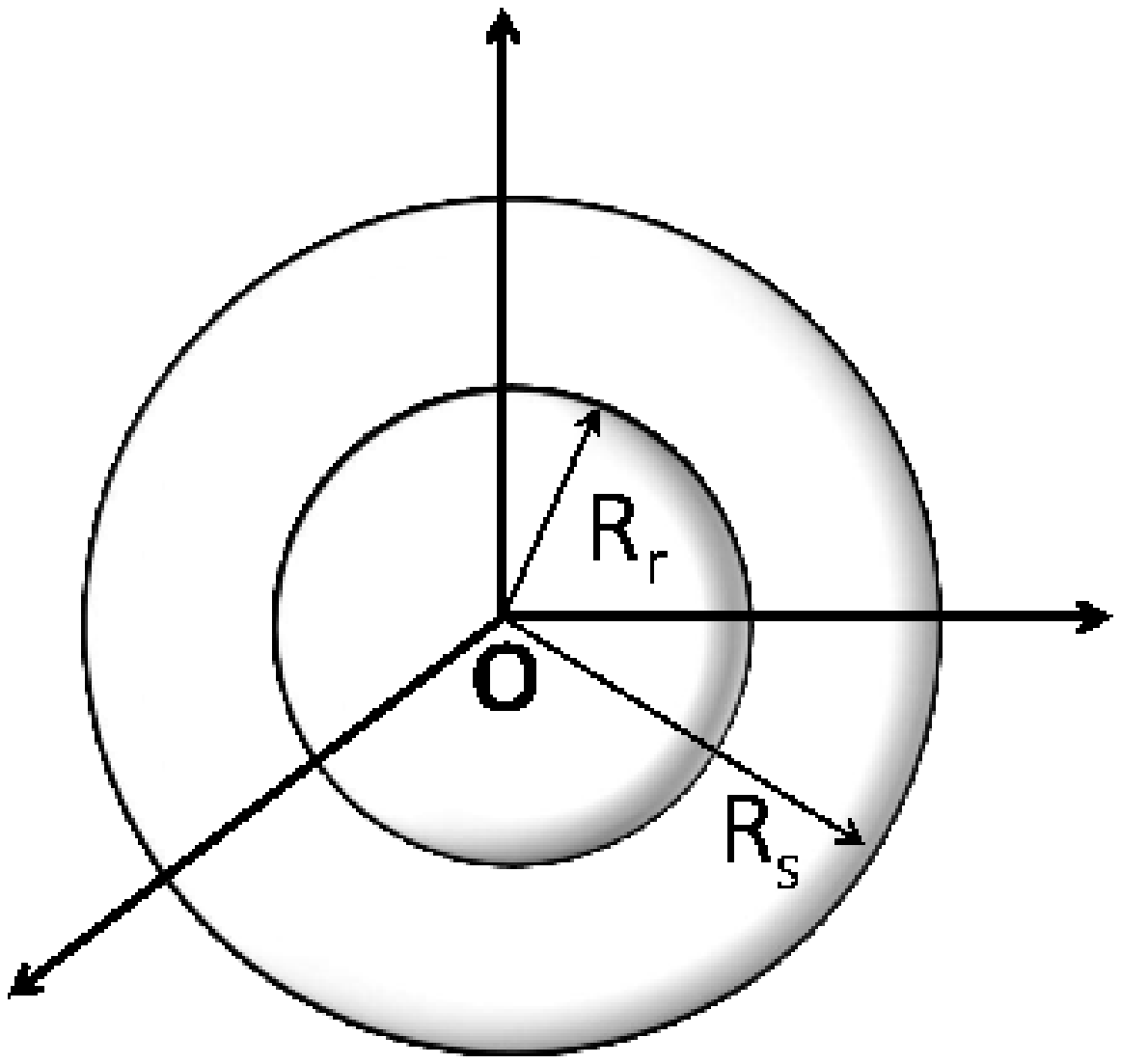}}\qquad
\subfigure[Overlapping]{
\label{fig:three}
\includegraphics[width=0.4\columnwidth]{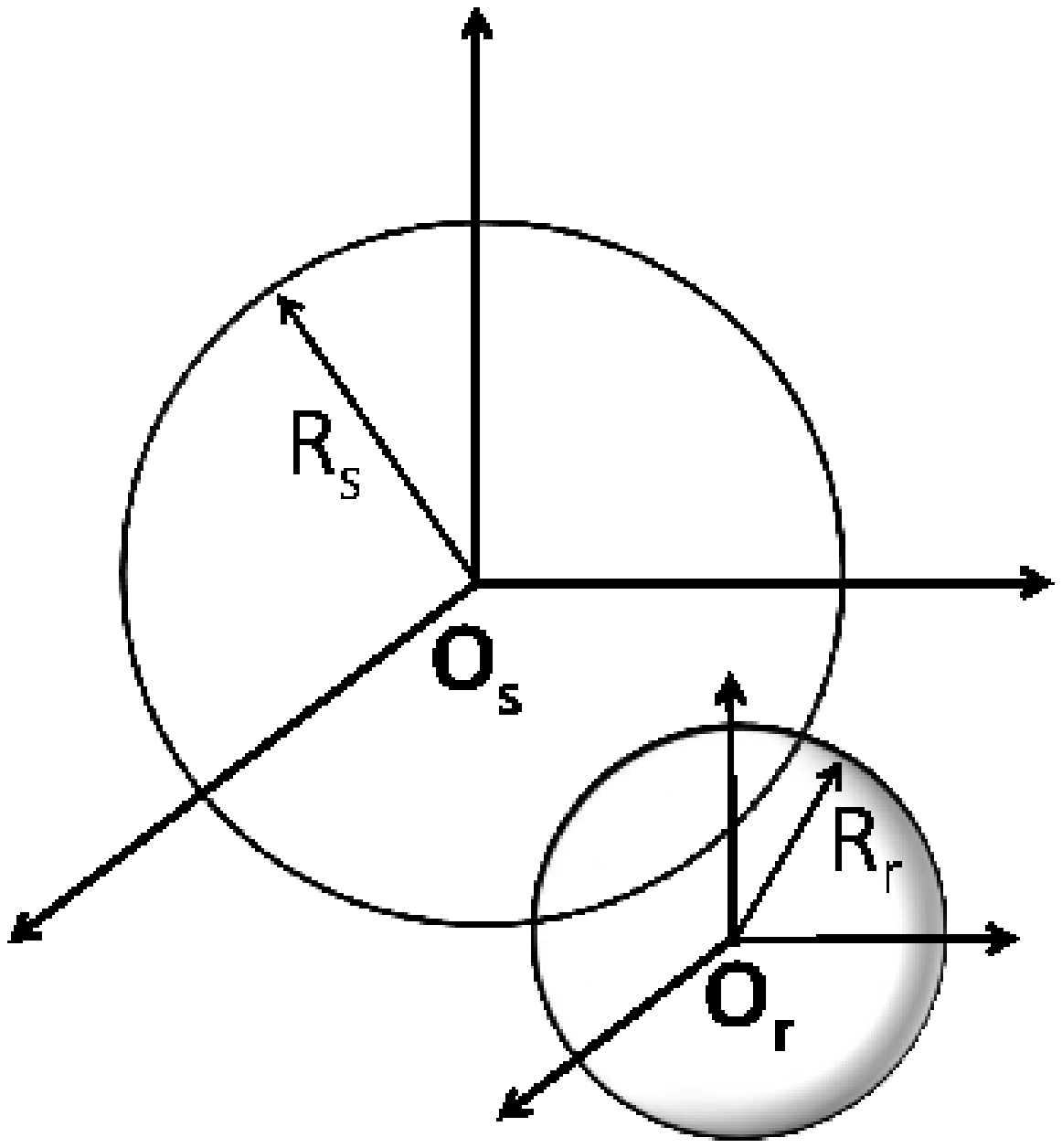}}%
\caption{Different configurations of the source region $\zeta$ and the receiver region $\eta$}
\label{fig:config}
\end{figure}

\noindent \textit{Comments:}\\
\begin{itemize}
\item Based on the above result  (\ref{eq:H}),  the RTF can be parameterized in terms of a spherical harmonic decomposition.  If  $\alpha^{nm}_{v\mu}(\textit{k})$, the weights/coefficients of this parameterization can be accurately captured, they can be used to derive the RTF between any two arbitrary points from a continuous spatial region where the source(s) lie and a continuous spatial region where the receiver(s) lie.\\
\item  To  generalize the RTF over an $N_{s}^{\text{th}}$ order source region $\zeta$, where $N_{s}=k_{\text{max}}eR_{s}/2$ and an $N_{r}^{\text{th}}$ order receiver region $\eta$, where $N_{r}=k_{\text{max}}eR_{r}/2,$ the above parameterization  requires a minimum of $(N_{r}+1)^2(N_{s}+1)^2$  unique coefficients of the form $\alpha^{nm}_{v\mu}(\textit{k})$. For example, when the maximum frequency of interest is $f_{\text{max}}1$ kHz and the source and receiver regions of interest are both spheres of radius $0.2$ m with $N_{s}=N_{r}=5$, a fixed number of $1296$ unique coefficients are required to calculate the RTF between any two arbitrary points  $\boldsymbol{x}$ and $\boldsymbol{y}$ from $\eta$ and $\zeta$ respectively. In broadband applications, the total coefficient count will increase with each frequency sample $f_{o}$ requiring an additional set of $(k_{o}eR_{r}/2 +1)^2 (k_{o}eR_{s}/2 +1)^2$ coefficients.\\
\item Due to the decomposition of direct and reverberant components, this parameterization supports any configuration of $\eta$ and $\zeta$. As shown in Fig.~\ref{fig:config} they can  be either completely separated  from each other with $\left\|\boldsymbol{R_{sr}}\right\| > (R_{r}+R_{s})$ (Fig.~\ref{fig:one}),  concentric with $\left\|\boldsymbol{R_{sr}}\right\|=0$ (Fig.~\ref{fig:two}) or  overlapping on each other with $\left\|\boldsymbol{R_{sr}}\right\| < (R_{r}+R_{s})$ (Fig.~\ref{fig:three}).  Therefore, (\ref{eq:H}) can be used to either  partially or fully characterize the room according to user requirements.\\
\item Taking all of the above properties into consideration, the proposed parameterization can be interpreted as a modal based solution to the wave equation in arbitrary room environments.  Compared to the classical mode solution to the RTF defined for rectangular rooms \cite{kuttruff2009room} , this parameterization has three main advantages. First, this method is applicable to any arbitrary room geometry. Second, practical room environments (having furniture etc.) with irregular boundary conditions are extremely difficult to be characterized by the classical solution, whereas the new method is valid for any arbitrary acoustic environment. Third, unlike the total mode count ($\frac{4\pi}{3}V(\frac{f}{c})$ where $V$ denotes room volume) of the classical model, that of the new model ($(N_{r}+1)^2(N_{s}+1)^2$ ) can be reduced by defining smaller $R_{s}$ and $R_{I}$ values to improve the computational efficiency. (This property is specially advantageous at the Schroeder frequency\cite{schroeder1996schroeder} where the classical model will require a very large mode count resulting in a Gaussian distribution.
\end{itemize}

\section{Estimation of room transfer function coefficients}
\label{sec:Coefs}

In this section, we present the procedure of estimating the RTF coefficients  $\alpha^{nm}_{v\mu}(\textit{k})$ of (\ref{eq:H}) for a pre-defined source region and a pre-defined receiver region. As explained earlier, $\alpha^{nm}_{v\mu}(\textit{k})$ represents the $v^{\text{th}}$ order and $\mu^{\text{th}}$ mode  reverberant field coefficient within $\eta$ caused by an unit amplitude  $n^{\text{th}}$ order and $m^{\text{th}}$ mode outgoing wavefront originated at $\zeta$ (\ref{eq:3DExtSound1_}). For each outgoing mode from $\zeta$, there will be $(N_{r}+1)^2$ number of unique coefficients describing the reverberant  field incident at $\eta$, and to generalize an $N_{s}^{\text{th}}$ order source region, a total number of at least $(N_{s}+1)^2(N_{r}+1)^2$  coefficients needs to be extracted.

It is important to note that, in practice, the following method does not require the physical production of unit amplitude outgoing modes from $\zeta$ and associated room response recordings,  but only requires the acquisition of  room response between a set of receivers distributed within $\eta$ and a set of loudspeakers distributed within $\zeta$, each transmitting  a unit amplitude signal. Furthermore, given the room characteristics remain stationary over time, these measurements can be obtained using a single microphone unit and a single loudspeaker unit. However, for the purpose of deriving this result, we will  discuss a method to generate unit amplitude modal wavefronts propagating outward from the source region, and a soundfield recording technique to extract the corresponding room responses. Theoretically, these processes are required to be repeated for a minimum of $(N_{s}+1)^2$ number of different cases, but their physical implementation will be  proven to be needless in sec.~\ref{sssec:HOmic}.

\subsection{Synthesis of a unit amplitude outgoing mode originated from the source region}
\label{ssec:Production}
Let us first consider the problem of producing a unit amplitude outgoing mode  from $\zeta$  with respect to  $\boldsymbol{O_{s}}$ (\ref{eq:3DExtSound1_}).   In order to account for all the significant outgoing modes from $\zeta$,  $n'$ and  $m'$ from (\ref{eq:3DExtSound1_}) has to be varied from $0 \text{ to } N_{s}$  and   from $-n \text{ to } n$ respectively. This results in a total number of $(N_{s}+1)^2$  distinct  soundfield production cases and corresponding weight vectors.

For each case, we propose a mode matching approach where the modal coefficients of the desired outgoing field (\ref{eq:beta2}) are matched with those of the outgoing wavefield produced by an array of  loudspeakers distributed within $\zeta$. Consider $L$ number of point sources arbitrarily distributed within  $\zeta$,  where the $\ell^{\text{th}}$ source $(\ell=1 \cdots L)$ is located at $\boldsymbol{y^{(s)}}_{\ell}=(y^{(s)}_{\ell},\theta^{(s)}_{y\ell},\phi^{(s)}_{y\ell})$ with respect to $\boldsymbol{O_{s}}$. The weighted sum of  loudspeaker outputs will produce an outgoing soundfield of the form (\ref{eq:3DExtSound_}) where $\beta_{nm}^{(s)}(\textit{k})$ is \cite{samarasinghe20123d}
\begin{equation} 
\label{eq:betaa1}
\beta_{nm}^{(s)}(\textit{k})=\sum_{\ell=1}^{L}w_{\ell}(\textit{k})i\textit{k}j_{n}(\textit{k}y^{(s)}_{\ell})Y^{*}_{nm}(\theta^{(s)}_{y\ell},\phi^{(s)}_{y\ell})
\end{equation}
\noindent with $w_{l}(k)$ representing the weights at each point source. Our objective is to derive loudspeaker weights that will produce (\ref{eq:beta2}), a unit amplitude outgoing wave of order $n'$ and mode $m'$.  

This can be achieved by equating (\ref{eq:beta2}) to (\ref{eq:betaa1}), which forms a set of linear equations of the form 
\begin{equation}
\label{eq:mat1}
\boldsymbol{T}\boldsymbol{w}^{n'm'}=\boldsymbol{\beta}^{n'm'}
\end{equation}
\noindent where 

\begin{equation}
\label{eq:tmat}
\boldsymbol{T} = i\textit{k}\begin{bmatrix}t_{00}(\textit{k},\boldsymbol{y^{(s)}}_{1} )& \cdots & \cdots & \cdots & t_{00}(\textit{k},\boldsymbol{y^{(s)}}_{L} )\\  \vdots &  \vdots &  \vdots  & \vdots &  \vdots \\t_{N_{s}N_{s}}(\textit{k},\boldsymbol{y^{(s)}}_{1} ) & \cdots & \cdots & \cdots &t_{N_{s}N_{s}}(\textit{k},\boldsymbol{y^{(s)}}_{L} )  \end{bmatrix} 
\end{equation}

\noindent is an $(N_{s}+1)^2\times L$ translation matrix with $t_{nm}(\textit{k},\boldsymbol{y^{(s)}}_{\ell} )=j_{n}(\textit{k}y^{(s)}_{\ell})Y^{*}_{nm}(\theta^{(s)}_{y\ell},\phi^{(s)}_{y\ell})$,    $\boldsymbol{w}^{n'm'}=[w^{n'm'}_{1}(\textit{k}) \\ \cdots w^{n'm'}_{\ell}(\textit{k}) \cdots w^{n'm'}_{L}(\textit{k})]^{T}$  
\noindent is an $L$ long  vector of loudspeaker weights  and  $\boldsymbol{\beta}^{n'm'}=[0 \cdots 0$ $1$ $0$ $\cdots$ $0]^{T}$  is an $(N_{s}+1)^2$ long vector of desired field coefficients where the $ (n'^2+n'+m'+1)^{\text{th}}$ element is $1$ while all others are zero.  Since $\boldsymbol{T}$ and $\boldsymbol{\beta}^{n'm'}$ are both known, the required  weights at each loudspeaker can be solved using
\begin{equation}
\label{eq:mat1Sol}
\boldsymbol{w}_{n'm'}=\boldsymbol{T}^{\dagger}\boldsymbol{\beta}^{n'm'}
\end{equation}
\noindent where $\boldsymbol{T}^{\dagger}$ denotes the pseudoinverse. To avoid spatial aliasing, 
\begin{equation}
\label{eq:L}
L\geq (N_{s}+1)^2 
\end{equation}
\noindent has to be satisfied with (\ref{eq:mat1Sol})  yielding the minimum energy weight solution. 

While (\ref{eq:mat1Sol}) provides a numerical solution to the required loudspeaker weights,  it is also important to decide upon a robust array geometry.  The  spherical geometry has been  widely used for  rendering and acquisition of spatial soundfields \cite{rafaely2005analysis,  abhayapala2002theory, meyer2002highly} however,  its performance in the above task can be predicted to be  less robust due to the Bessel functions present in $\boldsymbol{T}$. When $j_{n}(\textit{k}y^{(s)}_{\ell})$ of (\ref{eq:tmat}) approaches zero crossings, the condition number  of $\boldsymbol{T}$  increases\footnote{The $2-$norm condition number of a matrix $\boldsymbol{T}$ is defined by $\kappa_{2}(\boldsymbol{T})=\|\boldsymbol{T}\|_{2}\cdot \|\boldsymbol{T}^{\dagger}\|_{2}$ and for a well conditioned matrix, it will be close to $1$.}, and since the pseudoinverse of an  ill-conditioned matrix is often erroneous, those errors will be propagated to the weight solution in (\ref{eq:mat1Sol}). Similar issues were experienced in  spherical microphone arrays used for interior field recording  \cite{rafaely2005analysis}, which were later overcome by using   rigid arrays \cite{meyer2002highly} or variable radii arrays like the spherical shell array given in \cite{rafaely_08a} and  the dual spherical array given in \cite{balmages2007open}.   For the loudspeaker array of interest, a rigid geometry  requires the incorporation of scattering effects  and a dual spherical array requires twice the number of loudspeakers.   Therefore, in this paper, we opt for the  simplest geometry of choice, an open spherical shell. A spherical shell  array is equally distributed in the angular space while the distance to each loudspeaker $y^{(s)}_{\ell}$  randomly varies  (with a uniform distribution) between a virtual spherical shell of outer radius $R_{s}\text{ and an inner radius }R_{s}'$.  For in depth reasoning and additional solution to the open sphere inverse problem, the reader may refer to \cite{fazi2012nonuniqueness}"

An alternate approach for achieving robust loudspeaker arrays was recently introduced in \cite{poletti2012interior} for $2$D soundfields, and in \cite{samarasinghe20133d} for $3$D soundfields where the conventional circular/spherical arrays of monopole loudspeakers were replaced by those of higher order loudspeakers.  A $3$D higher order loudspeaker of order $D$ is capable of producing polar responses up to the $D^{\text{th}}$ order. This solution significantly reduces the minimum requirement of loudspeaker units by a factor of $1/(D+1)^2$ at the expense of increased complexity at each loudspeaker unit and therefore, it is more suitable for sound reproduction in large spatial areas.  Since the practical implementation of higher order loudspeakers are still in the design stage, the above approach is not used in this paper. However, the reader is encouraged to refer to \cite{poletti2012interior, samarasinghe20133d} for a detailed description of the array processing involved with sound rendering using higher order loudspeakers.

\subsection{Extracting the room response at the receiver region}
\label{ssec:recording}

Once the desired outgoing waves are synthesized at $\zeta$, the next step involves the extraction of  resulting room reflections incident at $\eta$. It is important to note that all recordings obtained at $\eta$ carry both direct and reflected wavefronts originated at $\zeta$, and since we are only parameterizing the reverberant field,  the direct field component at each sensor output must be  removed prior to further processing.

Furthermore, as the reverberant field of interest (\ref{eq:3DintSound}) is a source free incoming field, its extraction can be treated as an {\em interior field recording problem}. The conventional approach to record an $N_{r}^{\text{th}}$ order incoming spatial soundfield requires a minimum  of $(N_{r}+1)^2$ omnidirectional microphones equally distributed on a spherical surface enclosing the region of interest \cite{abhayapala2002theory}. However, this approach has been proven to be less robust due to the above mentioned "Bessel zero problem" and as explained earlier, alternate geometries were later proposed in \cite{meyer2002highly, balmages2007open, rafaely_08a} to overcome this issue.

A further improved solution to the interior field recording problem was recently introduced in \cite{samarasinghe20123d} where the omnidirectional microphones were replaced by higher order (HO) microphones. A higher order microphone of order $A$ is capable of recording an $A^{\text{th}}$ order spatial soundfield with respect to the microphone's local origin. Thus, the use of  $A^{\text{th}}$ order microphones in recording an $N_{r}^{\text{th}}$ order soundfield substantially reduces the minimum requirement of measurements by a factor of $1/(A+1)^2$ at the expense of added complexity at each microphone unit.  Compared to the conventional omnidirectional microphone array, this approach also showed a significant improvement in the condition number of the translation matrix,  which in turn  increased the array's robustness. Furthermore, unlike the higher order loudspeakers, the practical implementation of HO microphones are relatively simple and there exist a number of different designs that were successfully implemented in practice \cite{li_04b}. The "Eigenmike" is one such commercially available fourth order microphone with an active frequency range of $0-6.5$ kHz. 

Due to the aforementioned efficiency and availability of HO microphones, we propose an array of $Q$  identical $A^{\text{th}}$ order microphones to be employed in the coefficient extraction process. For each unit amplitude outgoing wavefield produced at $\zeta$, there will be an $N_{r}^{\text{th}}$ order  soundfield incident at $\eta$, and according to \cite{samarasinghe20123d}, the extraction of such a soundfield requires a minimum of  $Q \geq (N_{r}+1)^2/(A+1)^2$ HO microphone units  distributed in any arbitrary geometry enclosing  the region of interest.  The translation between the HO microphone outputs and the desired reverberant soundfield is based on a coefficient translation theorem developed in \cite{samarasinghe20123d}, which will be discussed in detail in sec.~\ref{sssec:arrayOfHO}.
\\
\subsubsection{Higher order microphone}
\label{sssec:HOmic}
Let us now briefly discuss the functionality of a $3$D HO microphone.  Consider the $q^{\text{th}} (q=1 \cdots Q)$ HO microphone located at $\boldsymbol{O_{q}}$ with  $\boldsymbol{R_{q}}=(R_{q},\theta_{q},\phi_{q})$ representing the vector connecting $\boldsymbol{O}$ to  $\boldsymbol{O_{q}}$. For numerical simplicity, we assume it is designed following  the open array geometry given in \cite{abhayapala2002theory}, where an $A^{\text{th}}$ order microphone is composed of an array of  $Q'\geq(A+1)^2$ number of omnidirectional microphones equally distributed along a virtual spherical surface  of radius $r=A\textit{c}/\pi e f_{\text{max}}$ where $f_{\text{max}}$ denotes the maximum frequency of interest in broadband operation.   The $q'^{\text{th}}$  $(q'=1 \cdots Q')$ microphone of this array   will be located at $\boldsymbol{r_{qq'}}=(r_{qq'},\theta_{qq'},\phi_{qq'})$ with respect to  $\boldsymbol{O_{q}}$ recording

\begin{equation} 
\label{eq:3DInt}
P^{(q)}_{q'}(\boldsymbol{r_{qq'}},\textit{k})=\sum_{a=0}^{A}\sum_{b=-a}^{a}\gamma_{ab}^{(q)}(\textit{k})j_{a}(\textit{k}r_{qq'})Y_{ab}(\theta_{qq'},\phi_{qq'})
\end{equation}

\noindent where $r_{qq'}=r$ for all $q'$ and $\gamma_{ab}^{(q)}(\textit{k})$ represents the  soundfield coefficients  with respect to  $\boldsymbol{O_{q}}$. Over the entire array, there will be a total of $Q'$ recordings of the above form, and based on the orthogonal property of spherical harmonics \cite{abhayapala2002theory}, they can be collectively combined to extract $\gamma_{ab}^{(q)}(\textit{k})$  using

\begin{equation} 
\label{eq:gamma}
\gamma_{ab}^{(q)}(\textit{k})=\frac{1}{j_{a}(\textit{k}r_{qq'})}\sum_{q'=1}^{Q'}P^{(q)}_{q'}(\boldsymbol{r_{qq'}},\textit{k})Y^{*}_{ab}(\theta_{qq'},\phi_{qq'}).
\end{equation}

When the above   microphone is used to record the room response caused by  the $n^{\text{th}}$ order and $m^{\text{th}}$ mode unit outgoing wavefield originated from $\zeta$, the  incident pressure at the $q'^{\text{th}}$ omnidirectional microphone  will be
\begin{equation} 
\label{eq:Pqq'}
P^{(q)}_{q'}(n,m,\boldsymbol{r_{qq'}},\textit{k})=\sum_{\ell=1}^{L}w^{nm}_{\ell}(\textit{k})H(\textit{k},\boldsymbol{x^{(q)}_{q'}},\boldsymbol{y_{\ell}})
\end{equation}
\noindent where  $H(\textit{k},\boldsymbol{x^{(q)}_{q'}},\boldsymbol{y_{\ell}})$ denotes the RTF between the omnidirectional receiver at $\boldsymbol{x^{(q)}_{q'}}=\boldsymbol{R_{q}}+\boldsymbol{r_{qq'}}$ and the weighted point source at $\boldsymbol{y_{\ell}}=\boldsymbol{y^{(s)}_{\ell}}+\boldsymbol{R_{sr}}$ with respect to $\boldsymbol{O}$. Substituting for (\ref{eq:gamma}) from (\ref{eq:Pqq'}) we derive   the corresponding outputs at the $q^{\text{th}}$ HO microphone as 
\begin{equation} 
\label{eq:gamma1}
\begin{split}
\gamma_{ab}^{(q,n,m)}(\textit{k})&=\sum_{\ell=1}^{L} w^{nm}_{\ell}(\textit{k})\\
& \underbrace{\frac{1}{j_{a}(\textit{k}r_{qq'})}\sum_{q'=1}^{Q'}H(\textit{k},\boldsymbol{x^{(q)}_{q'}},\boldsymbol{y_{\ell}})  Y^{*}_{ab}(\theta_{qq'},\phi_{qq'})}_{\widetilde{\gamma}_{ab}^{(q,\ell)}(\textit{k})}
\end{split}
\end{equation}
\noindent where $\widetilde{\gamma}_{ab}^{(q,\ell)}(\textit{k})$ denotes the incident soundfield coefficients at $\boldsymbol{O_{q}}$ caused by  a  unit amplitude loudspeaker located at $\boldsymbol{y_{\ell}}$. 

Consequently, if $\widetilde{\gamma}_{ab}^{(q,\ell)}(\textit{k})$, the room response between  the $\ell^{\text{th}}$ loudspeaker and the $q^{\text{th}}$ HO microphone can be recorded for all $L$ loudspeakers and all $Q$ HO microphones, then, $\gamma_{ab}^{(q,n,m)}(\textit{k})$ can be easily derived  using the linearity property given in (\ref{eq:gamma1}). This profound result significantly simplifies the coefficient extraction  process by completely eliminating the requirement for  (\ref{eq:mat1})'s physical implementation. Furthermore, all $(N_{s}+1)^2$ distinct cases of (\ref{eq:mat1}) can now be synthesized using the same set of $\widetilde{\gamma}_{ab}^{(q,\ell)}(\textit{k})$ measurements and appropriate numerical processing. 

\subsubsection{Removal of the direct soundfield}
As $\gamma_{ab}^{(q,n,m)}(\textit{k})$ are considered as the HO microphone outputs, it is essential to remove their direct path components prior to  further array processing. The coefficients  $\gamma_{ab}^{(q,n,m)}(\textit{k})$   can be decomposed into direct and reverberant field components as

\begin{equation} 
\label{eq:gamma2}
\gamma_{ab}^{(q,n,m)}(\textit{k})=\gamma_{ab(\text{dir})}^{(q,n,m)}(\textit{k})+\gamma_{ab(\text{rvb})}^{(q,n,m)}(\textit{k})
\end{equation}

where the direct field component  is known to be \cite{poletti2005three}

\begin{equation} 
\label{eq:gamma3}
\gamma_{ab(\text{dir})}^{(q,n,m)}(\textit{k})=\sum_{\ell=1}^{L}w^{(nm)}_{\ell}(\textit{k})i\textit{k}h_{a}(\textit{k}R_{q\ell})Y^{*}_{ab}(\theta_{q\ell},\phi_{q\ell})
\end{equation}

\noindent with $(R_{q\ell,},\theta_{q\ell},\phi_{q\ell})$ denoting the spherical coordinates of $\boldsymbol{R_{q\ell}}=\boldsymbol{y_{\ell}}-\boldsymbol{R_{q}}$. Therefore, once $\gamma_{ab}^{(q,n,m)}(\textit{k})$ are obtained,  (\ref{eq:gamma2}) and (\ref{eq:gamma3}) can be used to eliminate their direct field components.
\\
\subsubsection{Array of higher order microphones}
\label{sssec:arrayOfHO}
The final step in array processing involves the translation of $\gamma_{ab(\text{rvb})}^{(q,n,m)}(\textit{k})$  to the desired RTF coefficients $\alpha^{nm}_{v\mu}(\textit{k})$.  As mentioned earlier, this can be done following  the coefficient translation theorem introduced in \cite{samarasinghe20133d} as

\begin{equation} 
\label{eq:gammaAlpha}
\gamma_{ab(\text{rvb})}^{(q,n,m)}(\textit{k})=\sum_{v=0}^{N_{r}}\sum_{\mu=-v}^{v}\alpha^{nm}_{v\mu}(\textit{k})\hat{S}^{\mu b}_{va}(\boldsymbol{R}_{\boldsymbol{q}})
\end{equation}

\noindent where
\begin{eqnarray*} 
\label{eq:Shat}
\hat{S}^{\mu b}_{va}&=4\pi i^{a-v}\sum_{l=0}^{\infty} i^{l} (-1)^{2\mu-b}  j_{l}(\textit{k}R_{q})Y^{*}_{l(b-\mu)}(\theta_{q},\phi_{q}) \\
& \times \sqrt{(2v+1)(2a+1)(2l+1)/4\pi} W_{1} W_{2} \mbox{, with}
\end{eqnarray*}
\begin{align*}
W_{1} &= \begin{pmatrix}
    v & a & l \\
    0 & 0  &0 
\end{pmatrix} \mbox{ and}
&
W_{2} &= \begin{pmatrix}
    v & a & l \\
    \mu & -b  & (b-\mu)\end{pmatrix}
\end{align*}
\noindent representing Wigner $3-j$ symbols.
For all  $Q$ number of microphones, (\ref{eq:gammaAlpha}) can be interpreted in matrix form as,
\begin{equation} 
\label{eq:mat2}
\boldsymbol{\gamma} = \boldsymbol{T'}  \boldsymbol{\alpha}
\end{equation}
\noindent where   $\boldsymbol{\gamma}=[\gamma_{00}^{(1,n,m)},..\gamma_{AA}^{(1,n,m)},....\gamma_{00}^{(Q,n,m)},...\gamma_{AA}^{(Q)}]^{T}$  is a $Q(A+1)^{2}$ long vector,   $\boldsymbol{\alpha}=[\alpha_{00},......\alpha_{N_{r}N_{r}}]^{T}$ is a $(N_{r}+1)^2$ long vector, and \\  
\begin{equation}
\label{eq:Tmat}
\boldsymbol{T'} = \begin{bmatrix} \hat{S}^{00}_{00}(\boldsymbol{R}_{\boldsymbol{1}}) & \cdots & \cdots & \cdots & \hat{S}^{N_{r}0}_{N_{r}0}(\boldsymbol{R}_{\boldsymbol{1}})\\  \vdots &  \vdots &  \vdots  & \vdots &  \vdots \\ \hat{S}^{0A}_{0A}(\boldsymbol{R}_{\boldsymbol{Q}}) & \cdots & \cdots & \cdots & \hat{S}^{N_{r}A}_{N_{r}A}(\boldsymbol{R}_{\boldsymbol{Q}}) \end{bmatrix} 
\end{equation}
\noindent is a $Q(A+1)^{2} \times (N_{r}+1)^2$ matrix. As $\boldsymbol{T'}$ is known, and the local recordings in $\boldsymbol{\gamma}$ can be derived from (\ref{eq:gamma1}),  (\ref{eq:mat2}) can be solved to find the desired coefficients $\boldsymbol{\alpha}$, using
\begin{equation}
\label{eq:mat2S}
\boldsymbol{\alpha} = \boldsymbol{T'}^{\dagger}\boldsymbol{\gamma}.
\end{equation}

\noindent To avoid spatial aliasing, 
\begin{equation}
\label{eq:Q}
Q\geq (N_{r}+1)^2 / (A+1)^2
\end{equation}
\noindent has to be satisfied \cite{samarasinghe20123d} with (\ref{eq:mat2S})  yielding a least squares solution. 
\\
\subsection{Summary of the coefficient extraction process}

The coefficient extraction process involved with the RTF parameterization proposed over an $N_{s}^{\text{th}}$ order source region and an $N_{r}^{\text{th}}$ order receiver region is summarized as follows. Theoretically, this process requires $(N_{s}+1)^2$ number of distinct outgoing waves created at $\zeta$ and   the same number of reverberant field extractions  simultaneously carried out at $\eta$. Each case requires a minimum of  $(N_{s}+1)^2$ point sources or $(N_{s}+1)^2/(D+1)^2$  number of $D^{\text{th}}$ order loudspeakers at $\zeta$ to synthesize the outgoing field, and a minimum of $(N_{r}+1)^2$ omnidirectional microphones or $(N_{r}+1)^2/(A+1)^2$ number of $A^{\text{th}}$ order microphones to extract the reverberant field  at $\eta$.  Depending on the size and frequency content of the source and  receiver regions, the user can employ any combination of point sources, higher order sources, omnidirectional microphones and higher order microphones.  As explained in sections~\ref{ssec:Production} and \ref{ssec:recording},   this work illustrates one of the above combinations, an open spherical shell array of  point sources and an open spherical array of  HO microphones.

However, in practice, it is only required to extract the room response between each loudspeaker and each microphone from the above arrays, and by incorporating these measurements with the numerical computations given in (\ref{eq:mat1Sol}), (\ref{eq:gamma1}), (\ref{eq:gamma2}) and (\ref{eq:mat2S}), the desired RTF coefficients can be successfully derived. Moreover, given the room characteristics remain stationary over time, the above measurements can be obtained using a single microphone unit and a single loudspeaker unit moved along the respective arrays.

Practical limitations of the proposed measurement method arise with large $R_{s}$ and $R_{r}$ values at high frequencies due to the increased number of modal components required to describe the spatial soundfield of interest.  Common forms of  these limitations include the large requirement of microphone and loudspeaker numbers in non-stationary conditions,  increased demand for high computational power, and the design and implementation constraints involved with large spherical/shell geometries. Furthermore, the proposed use of HO microphones to record $\eta$  may not be practically feasible at present due to the lack of affordable   HO microphones that are commercially available.  The above constraints may  however be overcome by defining smaller source and receiver regions to suit the application of interest, and the HO microphone array can be easily replaced with  omnidirectional ones to reduce costs. In addition to the above constraints, in a real room, temperature (and to a lesser extent humidity) fluctuations can cause the room impulse responses to change, especially in the late reverberant tails and the higher frequency components.  However, the proposed approach is likely to be accurate at low frequencies and to the modeling the RTF components of low order reflections.

\subsection{Approximate parameterization error}
\label{ssec:error}
The total error involved with the proposed RTF parameterization has several components that will be encountered at different stages of the  parameterization process. The first component will appear at the loudspeaker array processing phase (\ref{eq:mat1}), in the forms of truncation error  ($N_{s}=\left\lceil \textit{k}eR_{s}/2\right\rceil$) in (\ref{eq:3DExtSound_}), and least squares error in (\ref{eq:mat1Sol}) related to the geometry and numbers of loudspeakers. The next component will occur at each HO microphone, again in the forms of truncation error  ($A=\left\lceil \textit{k}er/2\right\rceil$) in (\ref{eq:3DInt}), and Bessel-zero error in (\ref{eq:gamma}). The final  component will develop in the coefficient translation phase, once more in the forms of truncation error  ($N_{r}=\left\lceil \textit{k}eR_{r}/2\right\rceil$) in (\ref{eq:3DintSound}), and least squares error in (\ref{eq:mat2S}). A detailed  decomposition of each of the above components is of least interest in the current context, thus, we only study the total error. For computational simplicity, we define an approximate  error averaged over a finite number of design points from $\zeta$ and $\eta$ as

\begin{equation}
\label{eq:error}
E=\frac{\sum\limits_{g=1}^{G}\|\widetilde{H}(\boldsymbol{x}_{g},\boldsymbol{y}_{g},\textit{k})-H(\boldsymbol{x}_{g},\boldsymbol{y}_{g},\textit{k})\|}{\sum\limits_{g=1}^{G}\|\widetilde{H}(\boldsymbol{x}_{g},\boldsymbol{y}_{g},\textit{k})\|}
\end{equation}

\noindent where $G$ denotes the number of source and receiver point combinations being considered and $\boldsymbol{\widetilde{H}}$ denotes the existing RTF. 

\section{Simulations}
\label{sec:Sim}
In the following simulation examples, we illustrate the accuracy of the proposed RTF parameterization in broadband applications.  A $6\times5\times2.5$ m rectangular  room was considered as the reverberant environment with its center defined as the origin $\boldsymbol{O}$. The RTF was parameterized over a spherical receiver region $\eta$  of radius $R_{r}=0.4$ m centered about $\boldsymbol{O}$ and a spherical source region $\zeta$  of radius $R_{s}=0.4$ m centered about $\boldsymbol{O_{s}}$. The location of $\boldsymbol{O_{s}}$ was varied accordingly to simulate a non-overlapping and an overlapping configuration of $\zeta$  and $\eta$.  The design frequency range was assumed up to $f_{\text{max}}=1$ kHz producing a tenth order receiver region $(N_{r\text{(max)}}=10)$ and a tenth order source region $(N_{s\text{(max)}}=10)$.  For frequencies below $f_{\text{max}}$,  the truncations limits would drop, and therefore, when operating with varying frequencies, $N_{s}$  and $N_{r}$ were varied accordingly.

From (\ref{eq:L}), the synthesis of a unit amplitude outgoing wave from $\zeta$  required a minimum of $L=121$ point sources distributed in a preferred geometry. As explained in Sec. (\ref{ssec:Production}), we opted for a spherical shell geometry,  which required the sources to be equally distributed in the angular space  and randomly varied in the magnitude space. While \cite{Online} provided an approximate solution for the desired angular distribution, the distance to each  source $\left\|\boldsymbol{y}^{(s)}\right\|$ was randomly varied  (with uniform distribution) between a spherical shell of outer radius of $R_{s}=0.4$ m and an inner radius of $R_{s}'=0.3$ m. 

\begin{figure}[t]
  \centering
  \centerline{\includegraphics[width=\columnwidth]{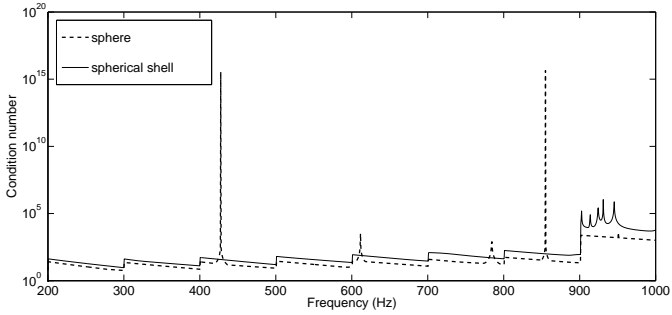}}
  \caption{Condition number variation of $\boldsymbol{T}$ in (\ref{eq:mat1}).} 
  \label{fig:cond}
\end{figure}

The reason behind selecting a spherical shell geometry over the conventional single sphere geometry was to improve the array robustness, and we validated this decision by comparing the condition number of the translation matrix $\kappa_{2}(\boldsymbol{T})$ related to both geometries. As shown in Fig.~\ref{fig:cond}, the condition number $\kappa_{2}(\boldsymbol{T})$ was plotted against frequency for a spherical shell geometry with the above parameters  and a  single sphere geometry of radius $R_{s}=0.4$ m. The expected ill-conditioning of the single sphere geometry is  very much evident with $\kappa_{2}(\boldsymbol{T})$ reaching a couple of large peaks at $f=420$ Hz and $f=850$ Hz. In contrast, $\kappa_{2}(\boldsymbol{T})$ of the spherical shell geometry gives much improved results by avoiding all of the above peaks. Therefore, we can conclude that a variation of $\left\|\boldsymbol{y}^{(s)}\right\|$ in $\boldsymbol{T}$ successfully overcomes the Bessel zero problem in solving (\ref{eq:mat1Sol}). The  sawtooth characteristic of the condition number variation may have caused by the frequency-dependent mode order ($N_{s}$).

\begin{figure}[t]
     \begin{center}
        \subfigure[]{%
            \label{fig:first}
            \includegraphics[height=0.35\columnwidth]{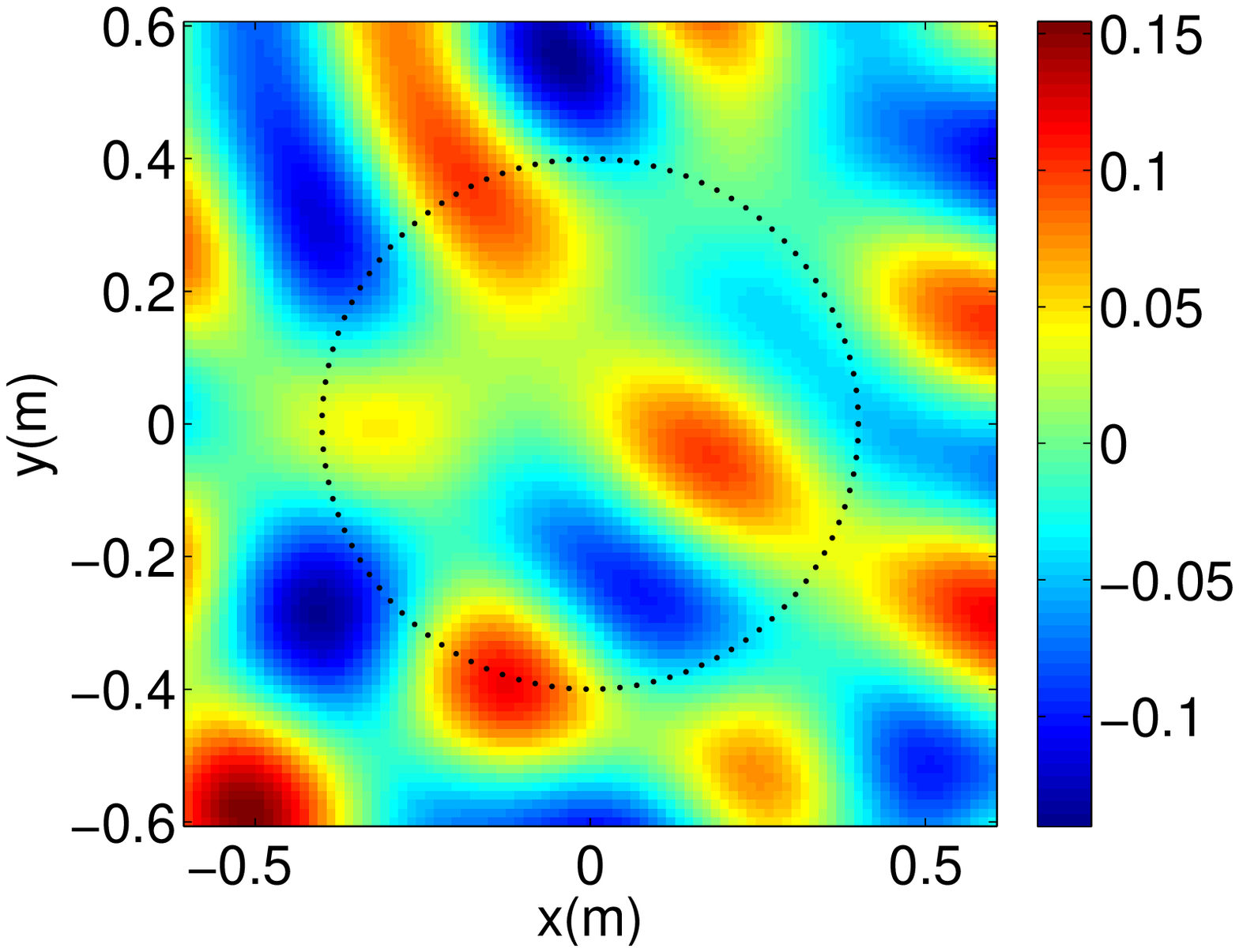}
        }%
        \subfigure[]{%
           \label{fig:second}
           \includegraphics[height=0.35\columnwidth]{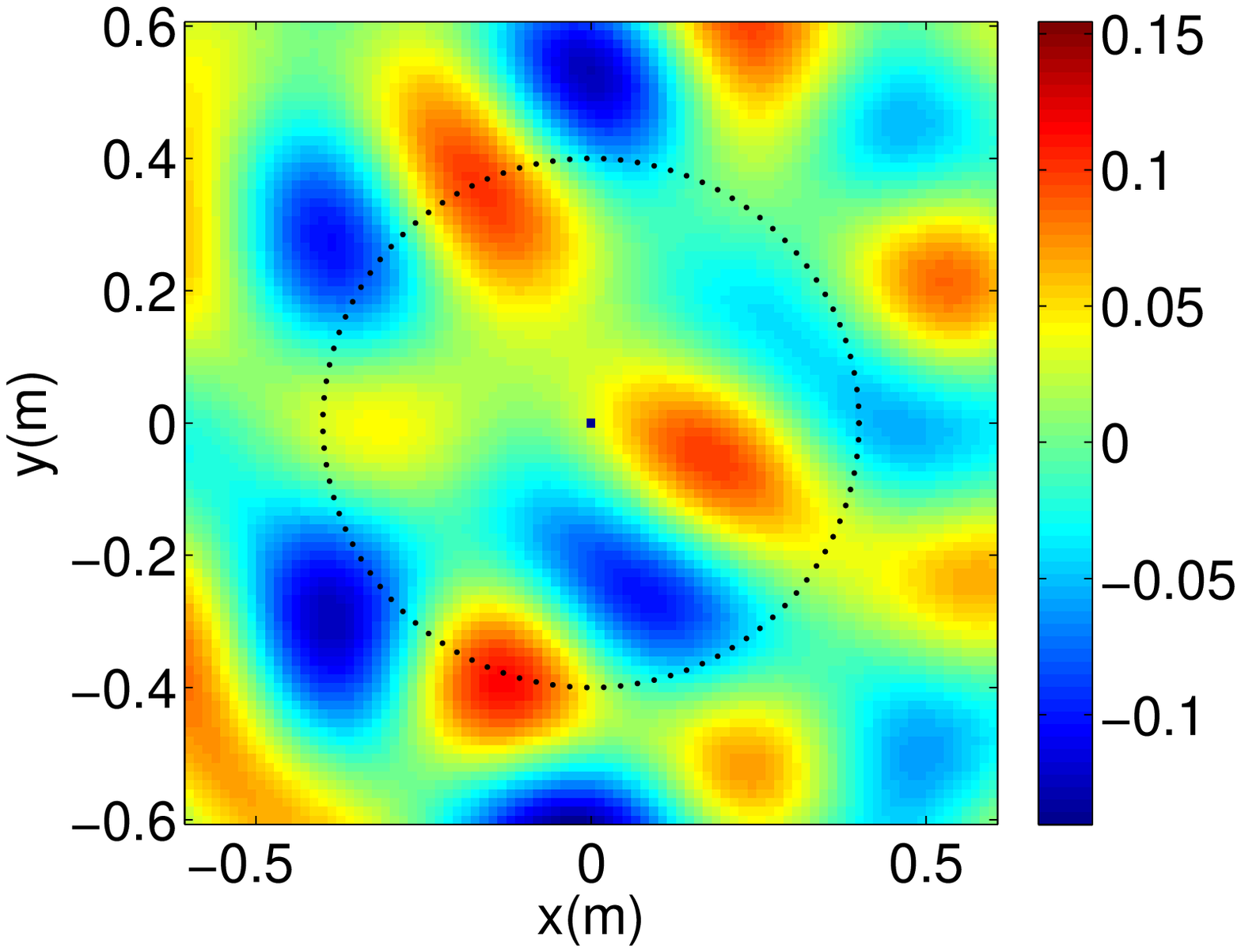}
        }
        \\
        \subfigure[]{%
            \label{fig:third}
            \includegraphics[height=0.35\columnwidth]{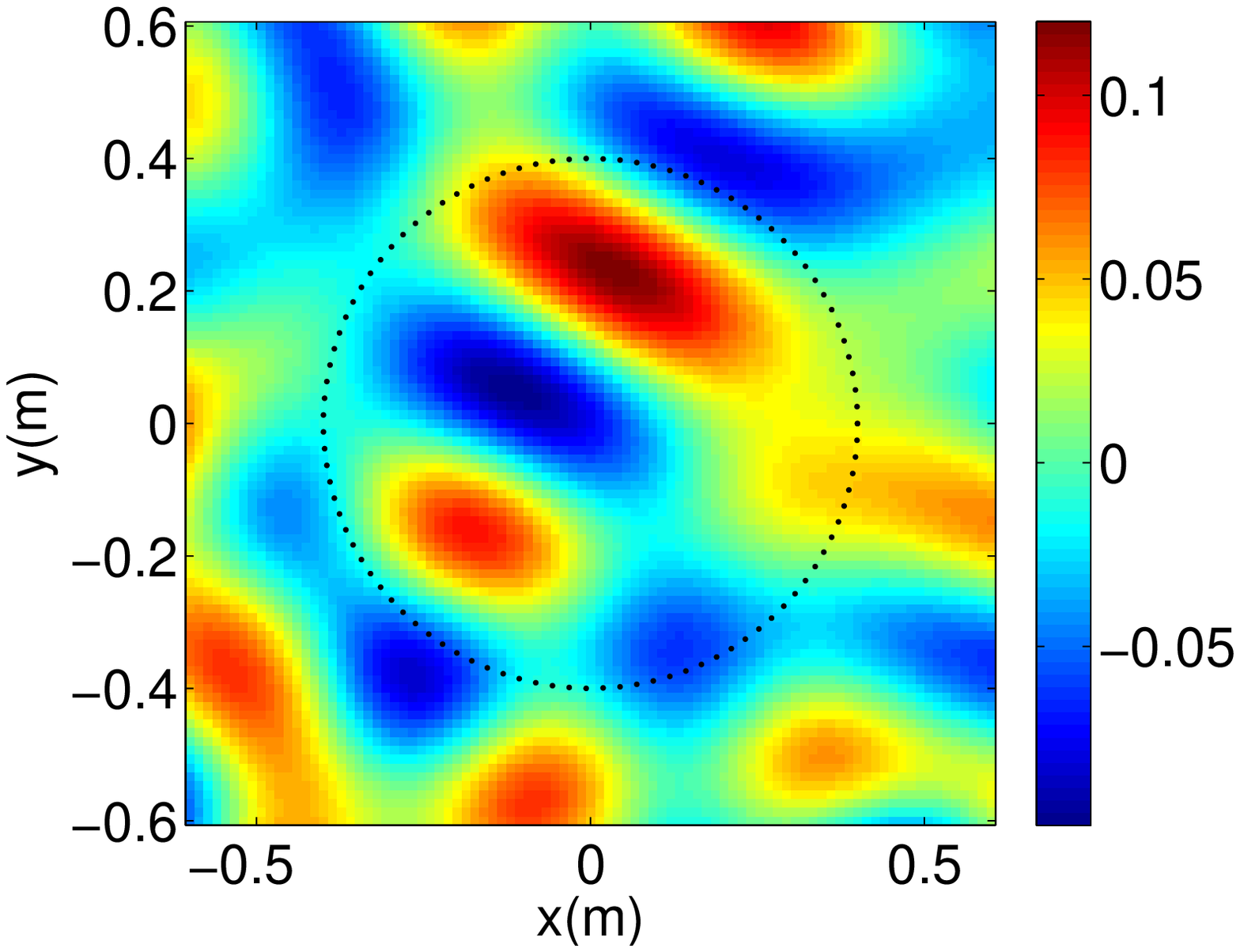}
        }%
        \subfigure[]{%
           \label{fig:fourth}
           \includegraphics[height=0.35\columnwidth]{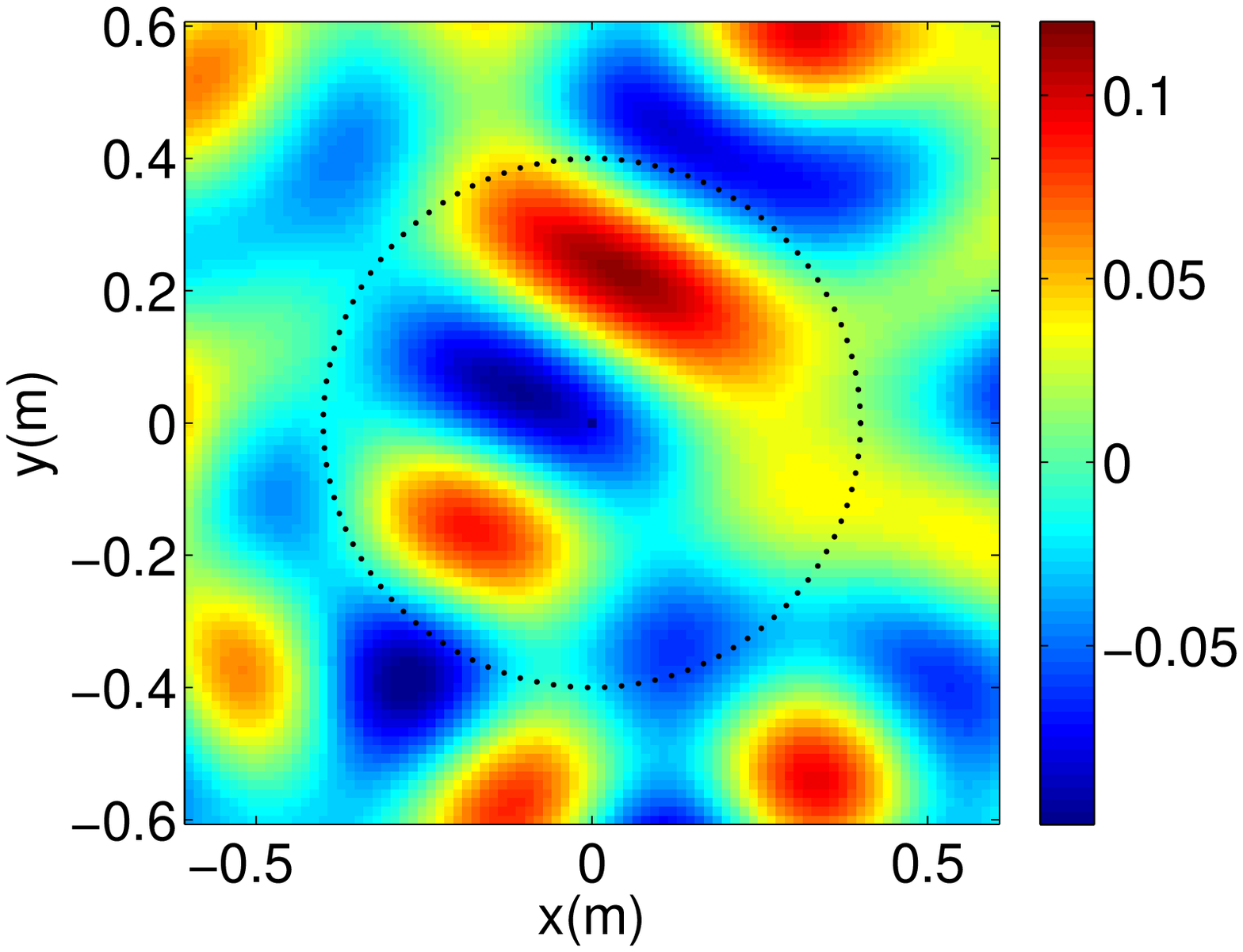}
        }%
    \end{center}
    \caption{%
        Actual and reconstructed  RTF  between a fixed source location $\boldsymbol{y}$  and all points in the receiver region for a non overlapping distribution of $\eta$ and $\zeta$ with $\boldsymbol{R_{sr}}=(1,\text{ } 1,\text{ }  0.5)$ m. (a) Actual and (b) reconstructed RTF for  $\boldsymbol{y}=(1.05, 1.05, 0.5707)$ m.    (c) Actual and (d) reconstructed RTF for  $\boldsymbol{y}=(1.15, 1.15, 0.6207)$ m.
     }%
   \label{fig:recordings}
\end{figure}

\begin{figure}[t]
     \begin{center}
        \subfigure[]{%
            \label{fig:first_src}
            \includegraphics[height=0.35\columnwidth]{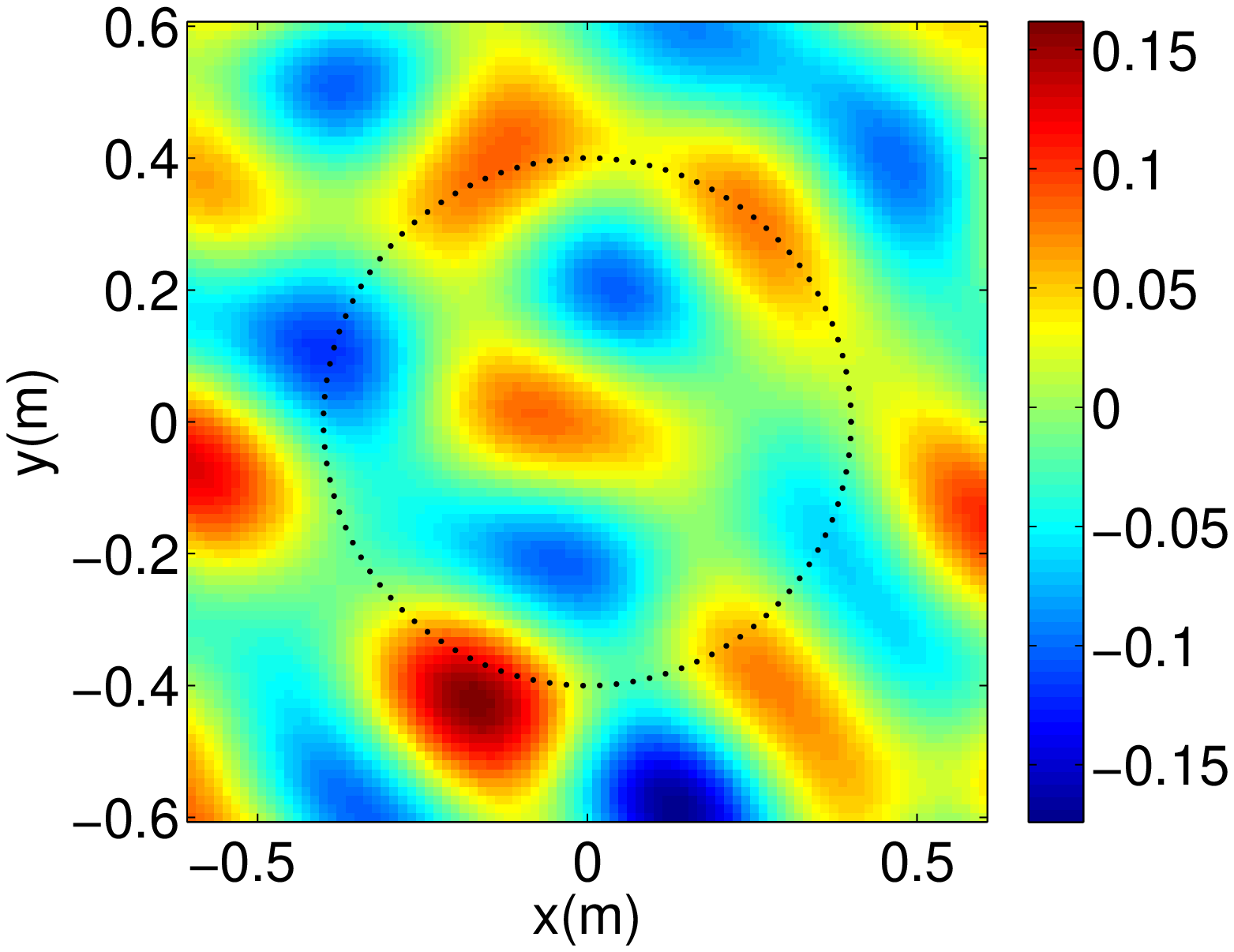}
        }%
        \subfigure[]{%
           \label{fig:second_src}
           \includegraphics[height=0.35\columnwidth]{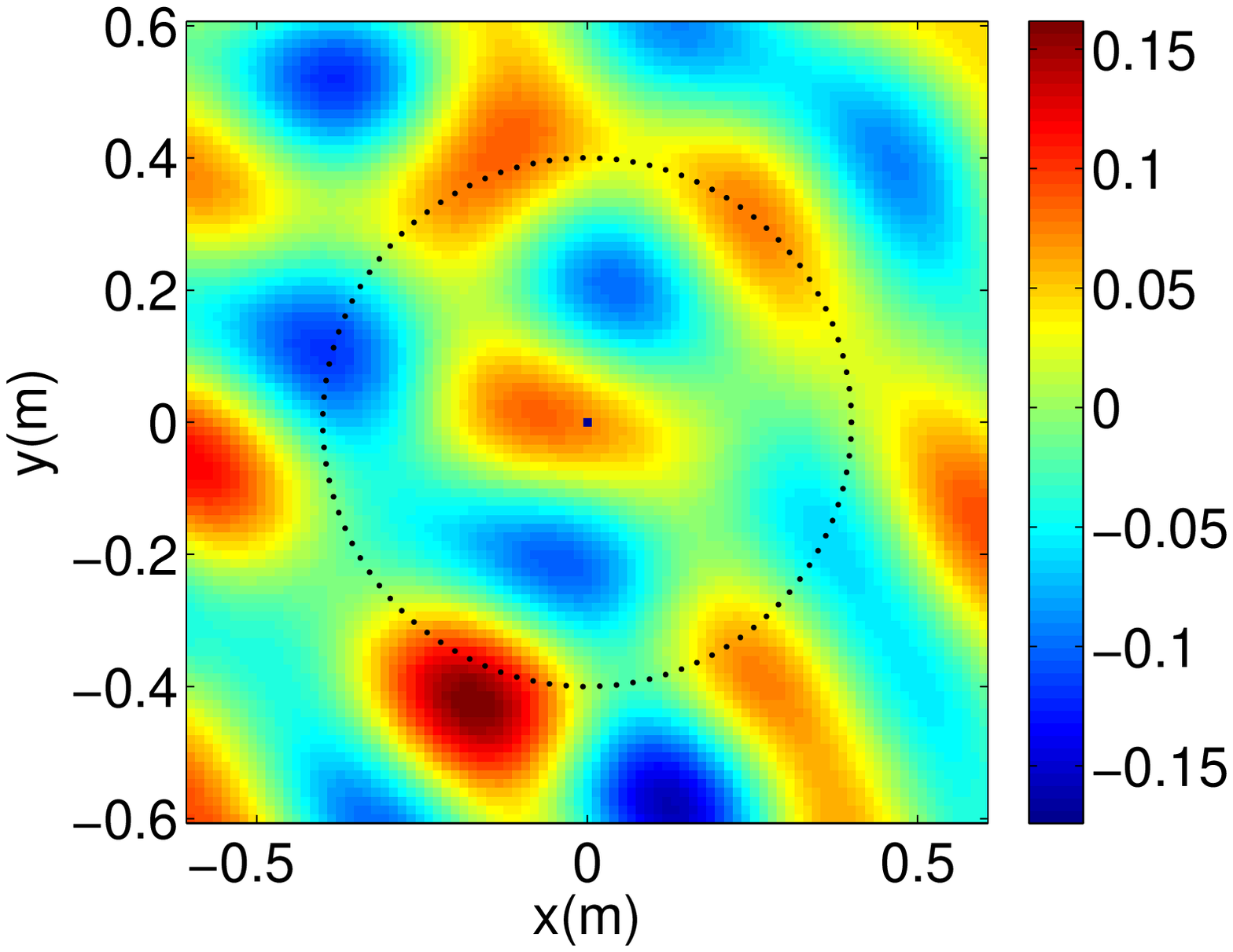}
        }
    \end{center}
    \caption{%
      (a)  Actual and (b) reconstructed  RTF  between a fixed receiver location at $\boldsymbol{y}=(0,0,0)$ m  and all points in the source region for a non overlapping distribution of $\eta$ and $\zeta$ with $\boldsymbol{R_{sr}}=(1,\text{ } 1,\text{ }  0.5)$ m.
     }%
   \label{fig:recordings2}
\end{figure}

\begin{figure}[]
  \centering
  \centerline{\includegraphics[width=\columnwidth]{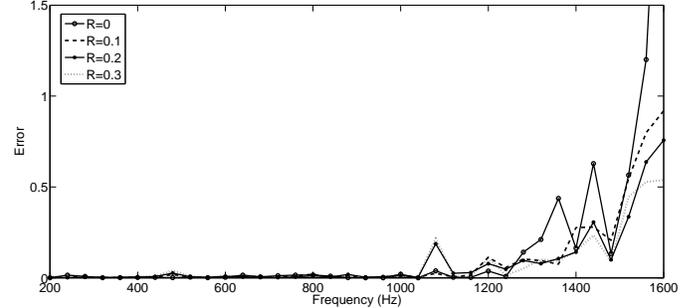}}
  \caption{ Approximate parameterization error (\ref{eq:error}) for $4$ different cases.  In each case,  the error was averaged over $G=7$ source and receiver point combinations, where each point was at $[(0,0,0),(-R,0,0),(R,0,0),(0,-R,0),(0,R,0),(0,0,-R),(0,0,R),]$ with respect to  $\boldsymbol{O_{s}}$ and $\boldsymbol{O}$ respectively.} 
  \label{fig:error}
\end{figure}%

Once the unit amplitude outgoing wavefields were synthesized at $\zeta$, it was required to extract the resulting reverberant fields incident at $\eta$.  For this purpose, we opted for a spherical array geometry of radius $R_{r}=0.4$ m enclosing $\eta$, and according to (\ref{eq:Q}), a minimum of $121$ omnidirectional microphones were required to avoid spatial aliasing. To improve the array robustness and to reduce the number of measurements, we replaced the omnidirectional microphones with third order ones $(A=3)$, which substantially reduced the minimum requirement of $Q$ to $(N_{r}+1)^2/(A+1)^2 \approx 9$. It is important to note that $Q$ was restricted to square numbers in order to facilitate the spatial distribution given in \cite{Online}, which provides an approximate solution to the equal division of a spherical surface.

However, in the simulations given below, we assumed the room acoustics to be stationary which in turn required only one point source and one third order microphone to measure the room response between $121\times 9$ combinations of $\boldsymbol{y_{\ell}}$ and $\boldsymbol{R_{q}}$.  The number of measurements can be further reduced by an approximate factor of $1/(D+1)^2$ if the point source was replaced by a higher order loudspeaker of order $D$. 

The actual room response measurements were simulated using the image-source method \cite{allen1979image} which defines the RTF between $\boldsymbol{x}$ and $\boldsymbol{y}$ in terms of
\begin{equation}
\label{eq:image}
H(\boldsymbol{x},\boldsymbol{y},\textit{k}) =h_{0}(\textit{k}\left\|\boldsymbol{x}-\boldsymbol{y}\right\|)+ \sum_{i=1}^{I}\zeta_{i}h_{0}(\textit{k}\left\|\boldsymbol{x}-\boldsymbol{y_{i}}\right\|)
\end{equation}
\noindent where $\boldsymbol{y_{i}}$ and $\zeta_{i}$  are the position and accumulated wall reflection coefficient of the $i^{\text{th}}$ image source.  In this paper, we considered image sources up to the second order  with wall reflection coefficients $[0.9$ $0.9$ $0.9$ $0.9$ $0.7$ $0.7]$.

We first looked at a non-overlapping distribution of $\eta$ and $\zeta$ by defining the vector from $\boldsymbol{O}$ to $\boldsymbol{O_{s}}$ as $\boldsymbol{R_{sr}}=(1\text{ } 1\text{ }  0.5)$ m. Once $\gamma i_{ab}^{(q,l)}(\textit{k})$ of (\ref{eq:gamma1}) were measured using the simulated environment given in (\ref{eq:image}), we calculated the loudspeaker weight vector $\boldsymbol{W_{nm}}$ for $(N_{s}+1)^2$ distinct cases accounting for all combinations of $n$ and $m$. Afterward, they were  incorporated in (\ref{eq:gamma1}) to calculate $\gamma_{ab}^{(q,n,m)}(\textit{k})$ which was later modified using (\ref{eq:gamma2}) and (\ref{eq:gamma3}) to derive  the HO microphone recordings of the reverberant field, $\gamma_{ab(\text{rvb})}^{(q,n,m)}(\textit{k})$. Finally, $\gamma_{ab(\text{rvb})}^{(q,n,m)}(\textit{k})$  were translated to the  desired RTF coefficients $(\alpha^{nm}_{v\mu}(\textit{k}))$, using the coefficient translation relationship given in (\ref{eq:gammaAlpha}). 

Even though the extracted RTF coefficients are capable of mapping each point in the source region to each point in the receiver region over $0-1$ kHz, it is not possible to plot them all at once. Therefore, we first demonstrate the array robustness to receiver variations by plotting the RTF between a particular point in the source region and all points in the receiver region for a single frequency. Next, we generated a similar plot for  a secondary source location to observe the array robustness to a slight variation in source positioning. In order to further validate this property,  we finally plotted the RTF between a particular point in the receiver region and all points in the source region, accounting for all possible source variations. Please note that all spatial plots  were constrained to a $2$D horizontal cross section with zero elevation for ease of presentation. 

Figures \ref{fig:first} and \ref{fig:second} show the actual and reconstructed RTF between a source at $\boldsymbol{y}=(1.05, 1.05, 0.5707)$ m and all points in the receiver region  for a frequency of $f=900$ Hz (the circle represents the receiver region). Similarly, Figs.~\ref{fig:third} and \ref{fig:fourth} show the corresponding results for a secondary source at  $\boldsymbol{y}=(1.15, 1.15, 0.6207)$ m.  The reconstructed results in both cases appear  almost error-less verifying the accuracy of the proposed parameterization and its robustness to receiver variations as well as a slight variation in source positioning.  Demonstrating  source variations in a larger scale, Figs.~\ref{fig:first_src} and \ref{fig:second_src}  show the actual and reconstructed  RTF between a fixed receiver at    $\boldsymbol{y}  = (0,0,0)$ m and  all points in the source region $\zeta$ (the circle represents the source region). As expected, the reconstructed field is almost the same as the actual one, verifying the proposed model's robustness to source variations. 

%

Analyzing the broadband performance of the proposed parameterization, we plotted the  reproduction error (\ref{eq:error}) of the recorded RTF against frequency over a range of $200-1700$ Hz. In order to average the error, we considered a sample set of $7$ points from the source region and $7$ points from the receiver region resulting in $G=7$ one-to-one combinations\footnote{One-to-one combinations meaning, the first source location paired with the first receiver location, the second source location paired with the second receiver location etc.}. The source and receiver points were located at $[(0,0,0),(-R,0,0),(R,0,0),(0,-R,0),(0,R,0),(0,0,-R), \\ (0,0,R),]$ with respect to  $\boldsymbol{O_{s}}$ and $\boldsymbol{O}$ respectively. Figure~\ref{fig:error} shows the results for different values of $R$. The error remains very low up to the maximum frequency of interest $1$ kHz, beyond which it slowly builds up.  The increasing error present from $1$ kHz onwards is  due to spatial aliasing in both reproduction and recording phases.  The low amplitude errors present within the active frequency range ($0.2-1$ kHz) are  possibly stemmed from the HO microphone simulations.  When a fixed $A^{\text{th}}$ order microphone with a maximum recordable frequency $f_{\text{max}}$ is employed to record low frequencies ($f<f_{\text{max}}$), $\gamma_{ab}$  mode will be successfully recorded only if $f<f_{\text{act}}^{(ab)}$ where $f_{\text{act}}^{(ab)}$ denotes the activation frequency of the $a^{\text{th}}$ order $b^{\text{th}}$ mode component of the soundfield of interest. In other words, at low frequencies, the soundfield modes that are actually present or activated are only up to the order $A' =\pi f e r /C$,  and all modes produced beyond $A'$ will  be erroneous due to the $1/j_{a}(\cdot) $ term in (\ref{eq:gamma}). In order to minimize these errors, the higher order modes can be discarded as they are simply absent in the actual soundfield. The same technique can be applied to the larger microphone array when calculating the  soundfield at $\eta$.  When the inactive modes are discarded as explained above, the matrix dimensions of (\ref{eq:Tmat}) will vary with varying frequency. An extensive study on this solution and the resulting improvement in array robustness to noise is presented  in \cite{samarasinghe2014wavefield} for the $2$D case.

\begin{figure}[]
     \begin{center}
        \subfigure[]{%
            \label{fig:first1}
            \includegraphics[height=0.35\columnwidth]{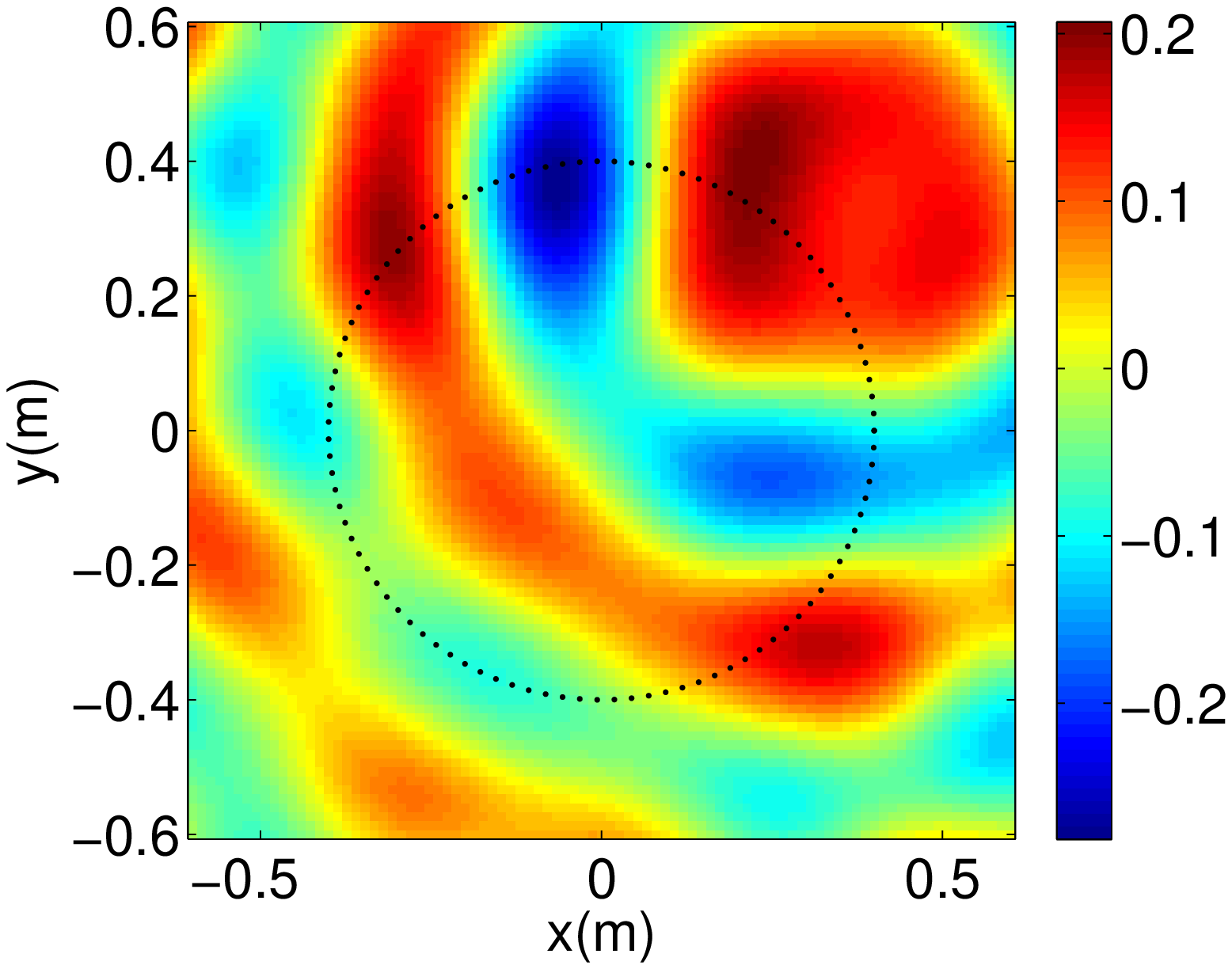}
        }%
        \subfigure[]{%
           \label{fig:second1}
           \includegraphics[height=0.35\columnwidth]{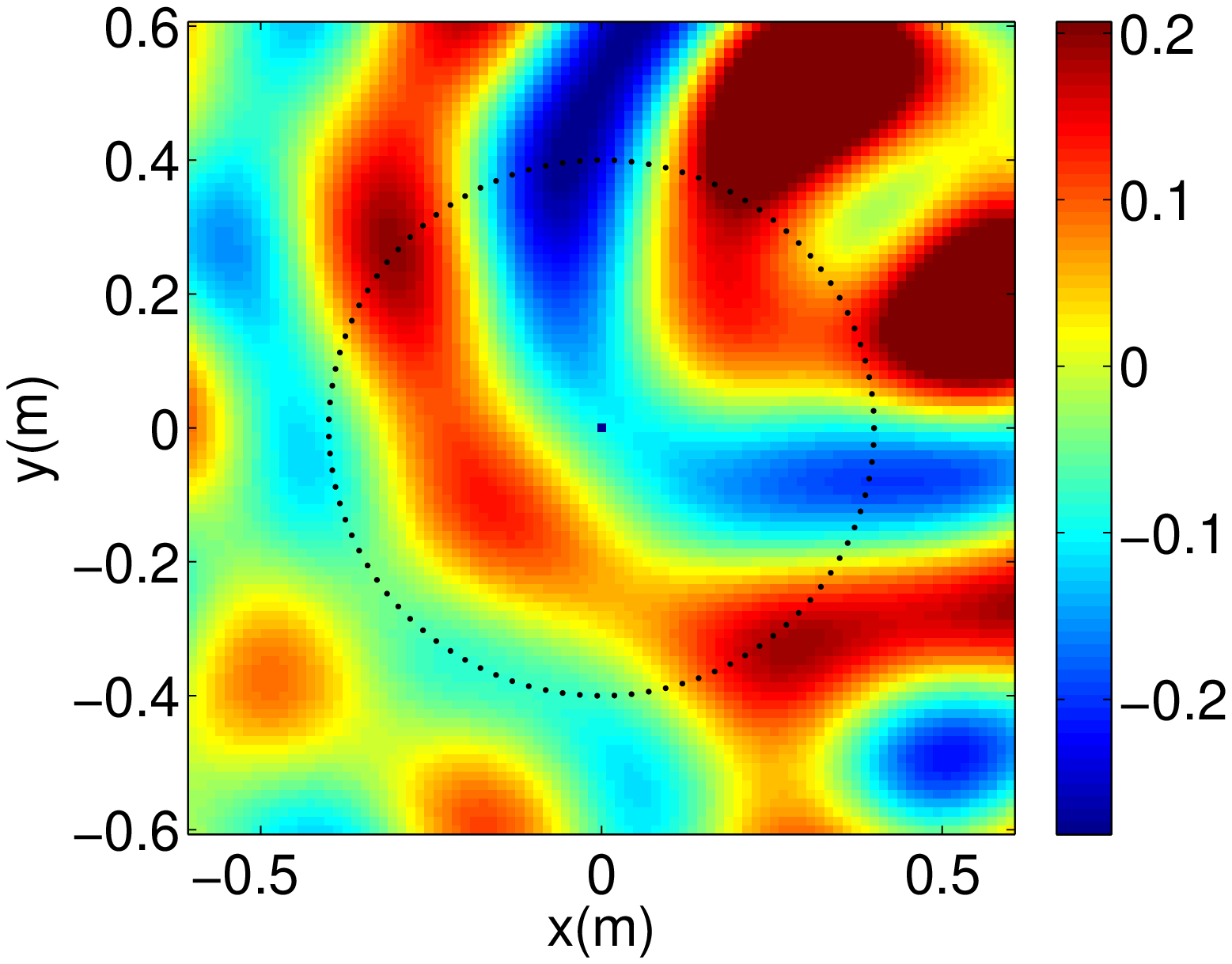}
        }
        \\
        \subfigure[]{%
            \label{fig:third1}
            \includegraphics[height=0.35\columnwidth]{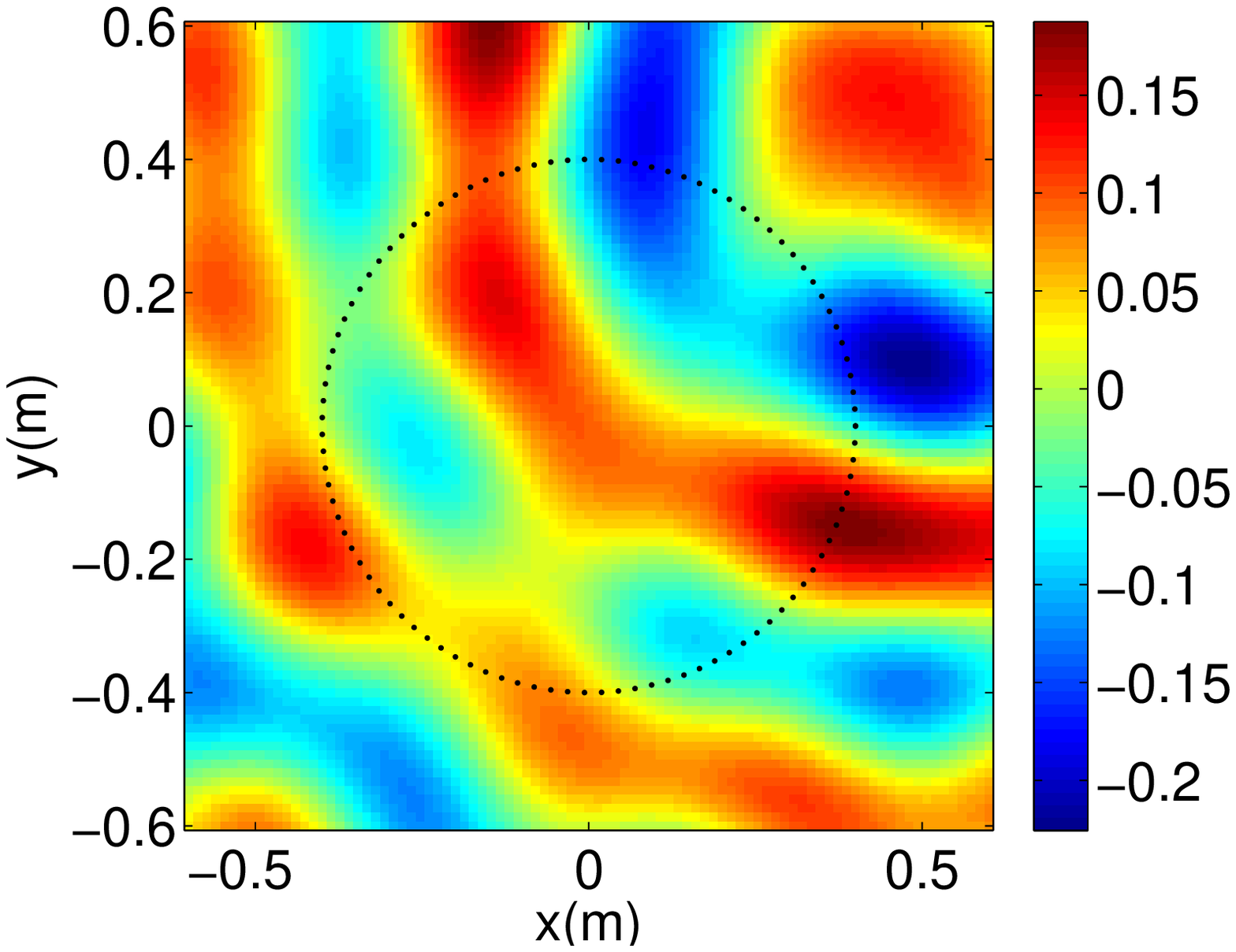}
        }%
        \subfigure[]{%
           \label{fig:fourth1}
           \includegraphics[height=0.35\columnwidth]{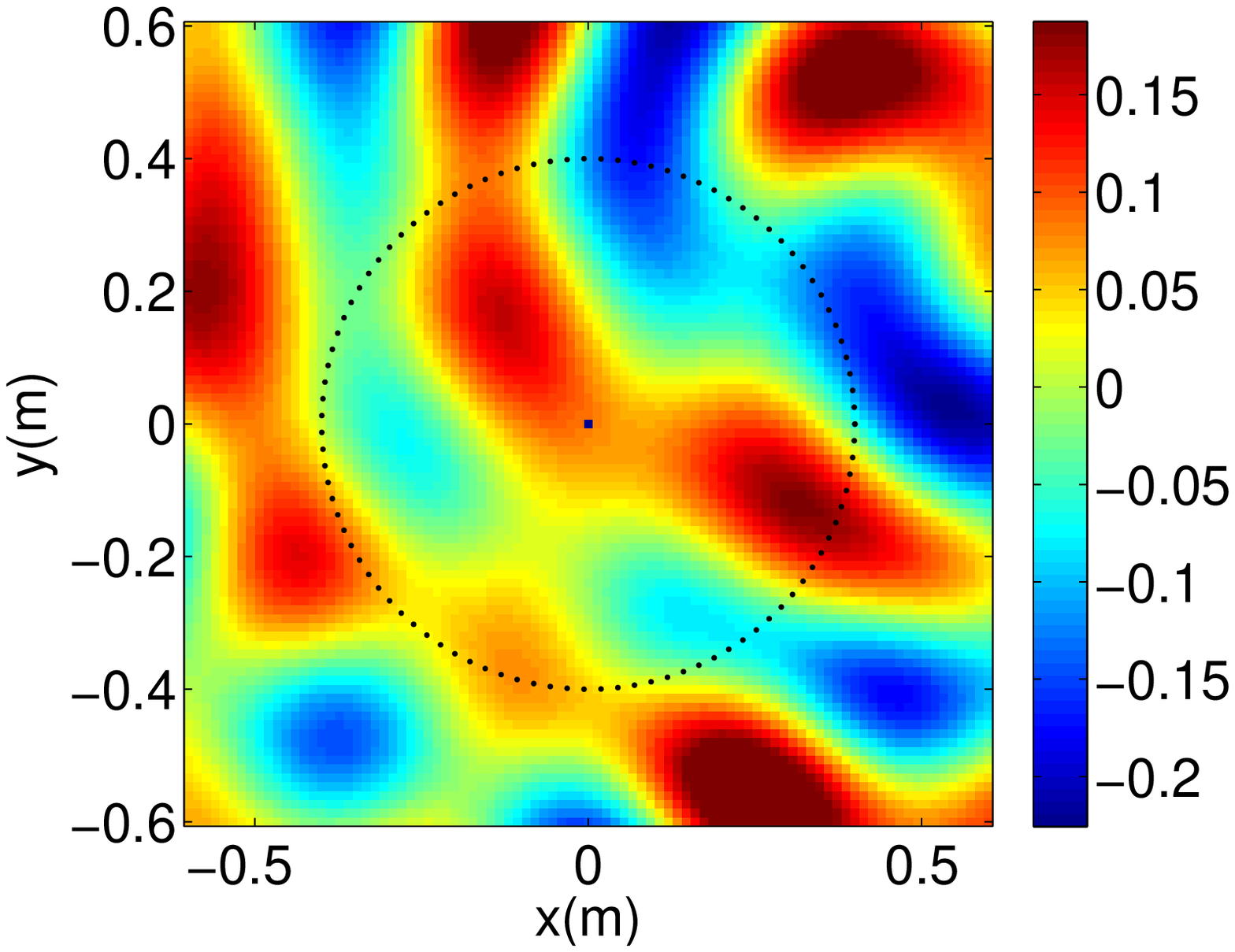}
        }%
    \end{center}
    \caption{%
        Actual and reconstructed  RTF  between a fixed source location $\boldsymbol{y}$  and all points in the receiver region  for an overlapping distribution of $\eta$ and $\zeta$ with  $\boldsymbol{R_{sr}}=(0.3\text{ } 0.3\text{ }  0.3)$ m. (a) Actual and (b) reconstructed RTF for  $\boldsymbol{y}=(0.35, 0.35, 0.3707)$ m.    (c) Actual and (d) reconstructed RTF for  $\boldsymbol{y}=(0.45, 0.45, 0.4207)$ m.
     }%
   \label{fig:recordings1}
\end{figure}

In order to verify the geometrical flexibility of the proposed parameterization, we repeated the same process for a different configuration of $\eta$ and $\zeta$. This was done  by re-defining the vector from $\boldsymbol{O}$ to $\boldsymbol{O_{s}}$ as $\boldsymbol{R_{sr}}=(0.3\text{ } 0.3\text{ }  0.3)$ m which resulted in $\eta$ and $\zeta$ to overlap on each other. However, all other design parameters were remained the same, which added no changes to the loudspeaker and microphone array parameters. Figures~\ref{fig:first1} and \ref{fig:second1} show the actual and reconstructed RTF between a source at $\boldsymbol{y}=(0.35, 0.35, 0.3707)$ m and all points in the receiver region  for a frequency of $f=900$ Hz (the circle represents the receiver region). Similarly, Figs.~\ref{fig:third1} and \ref{fig:fourth1} shows the corresponding results for a secondary source at  $\boldsymbol{y}=(0.45, 0.45, 0.4207)$ m.  The reconstructed results in both cases appear  almost error-less verifying the geometrical flexibility of the proposed parameterization. However, when  $\eta$ and $\zeta$ overlap on each other,   extra caution should be taken to avoid potential conflicts between the loudspeaker and microphone locations. Furthermore, when a loudspeaker is too near to a microphone, there will be potential errors stemmed from nearfield truncation \cite{koc1999error}.

\section{Conclusion}
\label{sec:Conclusion}
We have introduced a novel method to parameterize the three dimensional RTF between two arbitrary points from a sizeable spatial region where the source(s) lie and a sizeable spatial region where the receiver(s) lie. The modal based  parameterization only requires a finite number of RTF coefficients to describe an infinite number of RTFs between the two regions and therefore, it can also be used to characterize an entire room at once.  However, when an arbitrary shaped room is being measured for RTF parameterization, the corresponding microphone and loudspeaker array geometries may be altered to a similar geometry along (or close to) the room walls. We also presented a practical method of extracting the RTF coefficients which only requires a single loudspeaker and a single microphone, provided the room acoustics remain stationary.  This result   substantially simplifies the problem of room equalization by simplifying the RTF measurements. It also has the added advantage of providing robustness to both source variations as well as receiver variations.
\bibliographystyle{IEEEtran}
\bibliography{waspref}

\end{document}